\documentclass[preprint,10pt]{elsarticle}
\usepackage[margin=1.5cm]{geometry}% 
\usepackage{amssymb}
\usepackage{algorithm}
\usepackage{algpseudocode}
\usepackage{amsmath}
\usepackage{esint} 
\usepackage{bm}
\usepackage{color}
\usepackage{ulem}

\algdef{SE}[FUNCTION]{Function}{EndFunction}%
  {\algorithmicfunction\ }{\algorithmicend\ \algorithmicfunction}

\begin{document}

\begin{frontmatter}

\title{Simulations of Particle-Laden Flows with Large Dispersed-Phase Size Disparities Using Scalable Parallel Adaptive Methods
}

\author[inst1]{Linfeng Jiang}

\affiliation[inst1]{organization={Institute of Aerodynamics, RWTH Aachen University, Wüllnerstraße 5a, 52062 Aachen, Germany}  }

\author[inst2]{Enrico Calzavarini}
\affiliation[inst2]{organization={Université de Lille, Unit\'e de Mécanique de Lille - J. Boussinesq, UML ULR 7512, F59000 Lille, France}
            } 
\author[inst1,inst3]{Dominik Krug\corref{corresponding}}
\ead{d.krug@aia.rwth-aachen.de}
\affiliation[inst3]{organization={Physics of Fluids Group, Max Planck Center for Complex Fluid Dynamics, and J. M. Burgers Centre for Fluid Mechanics, University of Twente, P.O. Box 217, 7500 AE Enschede, The Netherlands}}

\cortext[corresponding]{Corresponding author}

\begin{abstract}
The numerical simulation of multiphase flows involving dispersed components with large scale disparities, such as the collisions between millimeter-sized bubbles and micron-sized mineral particles in flotation, poses a significant computational challenge. Accurately resolving the thin boundary layers of finite-size objects while tracking massive numbers of small particles within a large turbulent domain is often prohibitively expensive on uniform grids. To address this, we present a parallel scalable computational framework that couples the lattice Boltzmann method with the immersed boundary method on a dynamically adaptive octree grid.
A key algorithm is developed for the efficient parallel host-cell searching, which significantly accelerates the tracking of Lagrangian points on distributed unstructured grids. The accuracy and robustness of the code are rigorously validated against canonical benchmarks, including the flow induced by an oscillating cylinder and the sedimentation of a sphere. 
The framework is applied to the multiscale problem of bubble-particle collisions. In quiescent flow, the simulations accurately capture the hydrodynamic interception mechanism, reproducing the theoretical collision efficiency scaling law proportional to the square of the particle-to-bubble size ratio. Furthermore, the framework is applied to the simulation of fully resolved bubbles interacting with inertial point particles in homogeneous isotropic turbulence.

\end{abstract}

\end{frontmatter}

\section{Introduction}
\label{sec:introduction}

Fluids in nature and industry often contain suspensions of particles, bubbles, or droplets of various sizes. Collisions among these dispersed entities play a crucial role in many physical and environmental processes, i.e., the formation of raindrops through the coalescence of microscopic water droplets \cite{grabowski2013AFM}, sedimentation in rivers \cite{fan2023interaction}, fractional solidification of magmas \cite{PATOCKA2022117622}. Similar mechanisms are also essential in numerous industrial applications, including water treatment \cite{rubio2002WaterTreatment}, mineral recovery through flotation \cite{Nguyen2004}, and the separation of plastics for recycling \cite{kokkilicc2022plastic}. 
Numerical simulation provides a powerful tool to study such complex multiphase flow systems and to explore the dynamics of particle collisions under controlled conditions.
A central challenge in modeling such particle-laden flows is the accurate and efficient resolution of collisions between entities with large separations in scale, particularly within a turbulent background. For instance, in flotation, millimeter-sized bubbles interact with micron-sized mineral particles. These interactions involve high particle Reynolds numbers ($Re_p$) and thin boundary layers around the dispersed phase. Accurately capturing the two-way coupling between the particles and the surrounding fluid therefore necessitates interface-resolved simulation methods. In the context of the physical applications discussed in this work (e.g., flotation), bubbles are assumed to be fully contaminated by surfactants or impurities. Consequently, their interfaces are stiff, allowing them to be hydrodynamically modeled as non-deformable rigid spheres with a no-slip boundary condition.

The immersed boundary method (IBM) has emerged as a leading approach for simulating flows with finite-size particles~\cite{Uhlmann2005, verzicco2023ARF}. In the IBM framework, an interface is discretized by Lagrangian markers, and boundary conditions are enforced on the underlying Eulerian grid. This avoids the complexities of body-fitted meshing. However, this advantage comes at a steep price. To resolve the thin boundary layer around each particle, a fine grid resolution must be applied globally when using a uniform grid, causing the total number of grid points to scale severely with the Reynolds number~\cite{verzicco2023ARF}. The challenge is amplified in turbulent flows, where the grid must resolve both the particle boundary layers and the smallest dissipative eddies of the background flow. 
For instance, a finite-size particle at a high $Re_p$ often requires a large number of grid points across its diameter (e.g., $d_p/\Delta x = 100$) to accurately capture the boundary layer dynamics~\citep{johnson_flow_1999}. If this particle is marginally larger than the Kolmogorov scale (e.g., $d_p \approx 10\eta$), the required uniform grid spacing would be pushed to $\Delta x \approx 0.1\eta$. Since direct numerical simulation (DNS) of the background turbulence typically only requires $\Delta x \approx \eta$, employing a uniform grid implies that the background flow is over-resolved by an order of magnitude in each spatial direction. This leads to prohibitive computational costs and strongly highlights the necessity of using adaptive mesh refinement (AMR) to decouple the grid requirements of the particle boundary layer from those of the background turbulence.
By refining the mesh locally, fine grids can be concentrated around particle interfaces to resolve boundary layers, while coarser grids can efficiently represent the larger scales of the background flow. This approach not only dramatically reduces the total number of grid points but also maintains locally uniform grid patches, which is ideal for preserving the accuracy of the IBM interpolation schemes.

To solve the fluid dynamics on such adaptive grids, the lattice Boltzmann method (LBM) offers distinct advantages \cite{GUZIK2014461}. As a mesoscopic method with a simple stream-and-collide algorithm, the LBM is computationally efficient and its inherent data locality makes it exceptionally well-suited for massively parallel computing. Crucially, the LBM operates on a uniform Cartesian structure within each refinement level of the tree-based grid. This provides a natural and seamless compatibility with the IBM, avoiding the complex interpolations required when coupling traditional Navier-Stokes solvers with IBM on non-uniform grids. The combination of LBM-IBM on adaptive grids is thus a powerful and efficient framework for interface-resolved simulations.

While adaptive grids solve the resolution problem, their efficient parallelization presents challenges. For the Eulerian grid, tree-based structures are amenable to scalable partitioning using space-filling curves (SFCs), an approach leveraged by robust libraries like \texttt{p4est}\cite{burstedde2011p4est,isaac2015recursive}, which provides the foundation for our framework. Beyond the management of the Eulerian grid, however, coupling this framework with a large number of Lagrangian particles introduces another layer of algorithmic complexity. As particles advect across the domain, they frequently cross processor boundaries, necessitating efficient data migration. Furthermore, on a dynamically adapting grid, determining the specific grid cell (the ``host cell") that contains each Lagrangian point becomes a non-trivial and potentially expensive task. A fast and scalable parallel search algorithm for locating these host cells is therefore essential to the overall performance of the simulation.

This paper presents a scalable numerical method for simulating collisions of particles with large size disparities in turbulent flows, based on an LBM solver coupled with IBM on a dynamically adaptive octree grid. While our solver leverages the robust foundational data structures of existing libraries (e.g., \texttt{p4est} for octree management), the specific multi-scale coupling algorithms are developed specifically for this work. The main contributions of this paper are: 
(i) We developed a customized algebraic search algorithm and a two-tier MPI migration scheme tailored for \texttt{p4est}. This overcomes the lack of native tracking interfaces, enabling scalable communication for millions of particles across dynamically adapting grids. 
(ii) To prevent the Lagrangian search from being bottlenecked by massive turbulence-induced mesh refinement in the wake, we construct an independent, locally refined auxiliary grid forest exclusively for IBM markers, which significantly enhances the computational efficiency. 
(iii) By synergizing the tracking protocol and the auxiliary grid, we present a unified adaptive framework capable of simultaneously resolving finite-size bodies (via IBM) and tracking millions of one-way coupled sub-grid point particles.

This paper is organized as follows: In Section \ref{sec: numrics}, we detail the numerical methods, including the lattice Boltzmann method, the immersed boundary algorithm, and the governing equations for the dispersed phase. Section \ref{sec:parallel} extends these algorithms to the parallel, adaptive context. In Section \ref{sec:validation}, we present several validation cases that demonstrate the accuracy and scalability of our implementation. Finally, in Section \ref{sec:app}, we apply the validated framework to investigate the complex dynamics of bubble-particle collisions, highlighting the method's capability for studying complex multiphase systems.

\section{Numerical Methods} \label{sec: numrics}
\subsection{Lattice Boltzmann method}
\label{sec:LBM}
The fluid dynamics are solved using the LBM \cite{succi2001lattice,timm2016lattice}, which has been widely adopted in many open-source frameworks~\cite{calzavarini2019eulerian,heuveline2007openlb,Palabos2020}. The LBM describes the evolution of discrete-velocity distribution function, $f_i(\mathbf{x},t)$, at the position $\mathbf{x}$ and time $t$, where the index $i$ corresponds to a discrete velocity vector $\mathbf{c}_i$.  The governing lattice Boltzmann equation (LBE) is given by:
\begin{equation}
    f_i(\mathbf{x}+\mathbf{c}_i\Delta t, t+\Delta t) = f_i(\mathbf{x},t)+\lbrack \Omega_i(\mathbf{x},t) + S_i(\mathbf{x},t)\rbrack\Delta t\label{eq: LBE}
\end{equation}

where $\Delta t$ is time interval that it takes for particles $f_i$ to travel across one lattice space. The left-hand side represents the streaming of populations to neighboring lattice nodes, and the right-hand side, $\Omega_i(\mathbf{x},t)$, is the collision operator. The simplest collision process is modeled using the single-relaxation-time Bhatnagar-Gross-Krook (BGK) operator \cite{BGK1954model}:
\begin{equation}
\Omega_i(\mathbf{x},t) = -\frac{f_i-f_i^{eq}}{\tau}
\end{equation}\label{eq: collision_operator}

where $\tau$ is the relaxation time that controls the rate of approach to the equilibrium distribution function, $f_i^{eq}$. The equilibrium distribution, for a weakly compressible fluid, is a low-Mach number expansion of the Maxwell-Boltzmann distribution:
\begin{equation}
f_i^{eq}(\mathbf{x},t) = w_i\rho_f \left( 1+\frac{\mathbf{u}\cdot\mathbf{c}_i}{c_s^2}+\frac{(\mathbf{u}\cdot\mathbf{c}_i)^2}{2c_s^4}-\frac{\mathbf{u}\cdot\mathbf{u}}{2c_s^2}\right)
\end{equation}\label{eq: equilibrium_distribution}
Here, $w_i$ are weight coefficients specific to the chosen lattice stencil, and $c_s^2=(\Delta x/\Delta t)^2/3$ is the square of the lattice speed of sound, where $\Delta x$ is the lattice spacing. 
To incorporate the effects of an external body force $\mathbf{F}(\mathbf{x},t)$, which is essential for coupling the immersed boundary method, a discrete source term $S_i(\mathbf{x},t)$ must be incorporated into Eq. (\ref{eq: LBE}). We adopt the well-established forcing scheme proposed by Guo et al.~\cite{guo2002forcing} for calculating this source term.
The macroscopic fluid density $\rho_f$ and velocity $\mathbf{u}$  are recovered by taking moments of the distribution functions:
\begin{eqnarray}
\rho_f(\mathbf{x},t) &=& \sum_i f_i(\mathbf{x},t) \label{eq: cal_rho}\\
\mathbf{u}(\mathbf{x},t) &=& \frac{1}{\rho_f} \sum_i\mathbf{c}_i f_i(\mathbf{x},t) + \frac{\mathbf{F}\Delta t}{2\rho_f}\label{eq: cal_velocity} 
\end{eqnarray}
The fluid pressure is related to the density by $p=c_s^2\rho_f$ and the kinematic viscosity $\nu$ is determined by the relaxation time:
\begin{equation}
\nu = c_s^2(\frac{1}{\omega}- \frac{1}{2})\Delta t \label{eq: viscosity}
\end{equation}
where $\omega=1/\tau$ is the relaxation rate.
In this work, we employ the D2Q9 lattice for two-dimensional simulations and the D3Q19 lattice for three-dimensional simulations~\cite{qian1992lattice}.
The terminology D$k$Q$n$ refers to $k$ dimensions and number $n$ of discrete velocity vectors. 

\begin{figure}
  \centerline{\includegraphics[width=1.0\linewidth]{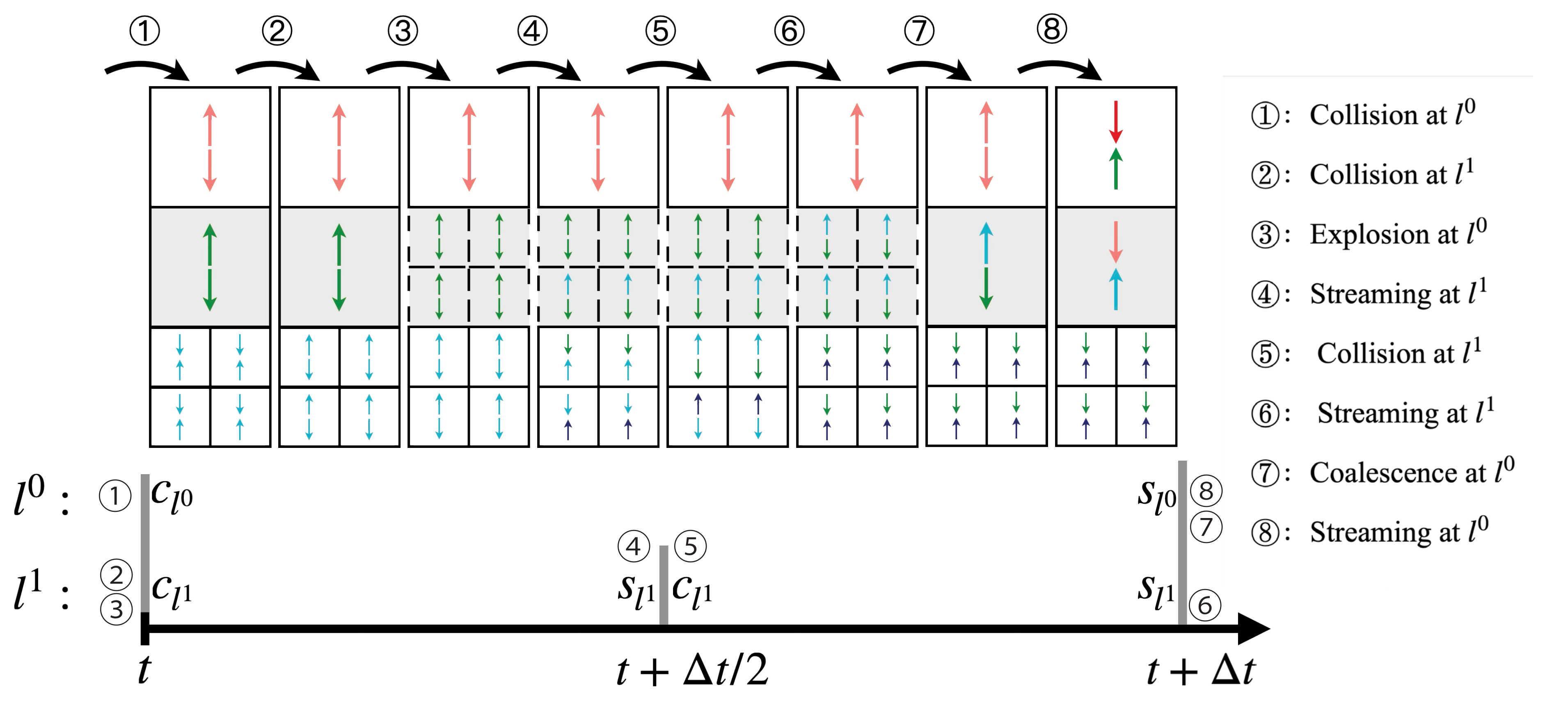}}
  \caption{Schematic illustration of one collision-streaming loop on the coarse grid (level $l^0$) and corresponding two collision-streaming loops on the fine grid (level $l^1$). The gray grid indicates the grid at the refinement border, which hosts the virtual cell. $c_{l^i}$ indicates the collision step at grid level $l^i$ and $s_{l^i}$ indicates the streaming step at grid level $l^i$. }
  \label{fig:lbm_sketch}
\end{figure}

\subsubsection{LBM on adaptive grids}
Generalizing the LBM to adaptive grids requires a consistent treatment of interfaces between different refinement levels. Two approaches are commonly used: (i) employing a constant time step across all levels, which alters the local Courant-Friedrichs-Lewy (CFL) number~\cite{Fakhari2014PRE,Fakhari2015CF}, and (ii) maintaining a constant ratio between the grid spacing and the time step, which introduces space-time adaptivity (often referred to as acoustic scaling~\cite{Lagrava2012JCP,thorimbert2022local} because the speed of sound remains constant). The latter can be achieved with temporal interpolation techniques~\cite{geier2009bubble} or volumetric formulations~\cite{chen1998volumetric, rohde2006generic}.
Approach (ii) is selected because it maintains a uniform acoustic speed across all levels, preventing spurious wave reflections and numerical instabilities at grid interfaces. Furthermore, since the global minimum time step is dictated by the finest level, Approach (ii) avoids forcing coarse grids to update at unnecessarily small time increments, thereby optimizing computational effort on coarse levels.
In this work, we adopt the volumetric, cell-centered formulation (Approach (ii)) proposed by Rohde et al.\cite{rohde2006generic}, which has been successfully implemented in High Performance Computing (HPC) codes like PEANO \cite{neumann2013dynamic} and waLBerla\cite{schornbaum2016massively}. 
Under acoustic scaling, the time step $\Delta t$ at a given refinement level $l$ is proportional to the grid spacing $\Delta x_l$~\cite{schonherr2011multi}. For the 2:1 refinement ratio used here, a coarse cell at level $l$ evolves with a time step $\Delta t_l$, while its child cells at level $l+1$ perform two complete collision-and-stream cycles using a time step $\Delta t_{l+1} = \Delta t_l/2$.

Communication across refinement boundaries is managed through a layer of virtual cells embedded in the coarse cell adjacent to the interface (see grey region in Fig.~\ref{fig:lbm_sketch}). The procedure to synchronize the grid over one coarse time step $\Delta t_l$ is as follows (see Fig.~\ref{fig:lbm_sketch}):
(i) \textbf{Collision:} Perform the collision step in both coarse and fine level of grids (step 1 and 2 in Fig.~\ref{fig:lbm_sketch}).
(ii) \textbf{Expansion:} Post-collision populations from the coarse cell ($f_i^c$) are directly assigned to the overlapping virtual sub-cells ($f_i^v$), such that $f_i^v = f_i^c$. This equal interpolation is physically consistent, as the distribution functions represent densities (step 3 in Fig.~\ref{fig:lbm_sketch}).
(iii) \textbf{Fine Grid Sub-cycles:} Two full collision-and-stream sub-cycles are performed on the fine grid. During streaming, populations are exchanged between real fine cells and the populated virtual cells (step 4-6 in Fig.~\ref{fig:lbm_sketch}).
(iv) \textbf{Coalescence:} The final post-collision populations in the virtual cells are averaged and transferred back to the corresponding coarse grid cell. For a $D$-dimensional problem, these values are simply transferred by evaluating $f_i^c = \frac{1}{2^D}\sum f_i^v$ (e.g., $f_i^c = \frac{1}{4}\sum f_i^v$ in 2D), ensuring strict mass and momentum conservation (step 7 in Fig.~\ref{fig:lbm_sketch}).
(v) \textbf{Coarse Grid Stream:} The coarse grid performs its streaming step using the updated distribution functions from the coalescence step (step 8 in Fig.~\ref{fig:lbm_sketch}).

To maintain physical consistency across refinement levels, quantities must be properly rescaled. With acoustic scaling, we have $\Delta x_c = 2\Delta x_f$ and $\Delta t_c = 2\Delta t_f$ (let $f$ and $c$ represent the quantities in fine and coarse grids). The speed of sound $c_s = \Delta x/(\sqrt{3}\Delta t)$ remains constant across all levels, ensuring that velocity in lattice units is continuous ($u_c=u_f$). The condition that the Reynolds number for a given flow should be grid independent, reads:
\begin{equation}
    Re_c = u_cL_c/\nu_c =Re_f = u_fL_f/\nu_f,
\end{equation}
this leads to the scaling relation of kinematic viscosity in lattice units across the grids: $\nu_f=\frac{\Delta x_c}{\Delta x_f}\nu_c$. Combining with Eq.~(\ref{eq: viscosity}), these imply that the relaxation rate $\omega$ is scaled with each grid level \cite{Lagrava2012JCP}:
\begin{equation}
    \omega_f = \frac{2\Delta x_f\omega_c}{\Delta x_f\omega_c+2\Delta x_c-\Delta x_c\omega_c} = \frac{2\omega_c}{4-\omega_c}
\end{equation}
A body force term, represented by a lattice acceleration $a$, must also be scaled to represent the same physical acceleration. This yields the relation:
\begin{equation}
    a_{c}\frac{\Delta x_c}{\Delta t_c^2}=a_{f}\frac{\Delta x_f}{\Delta t_f^2}  \Rightarrow a_{c} = a_{f}\frac{\Delta t_c}{\Delta t_f}\Rightarrow \quad a_c = 2a_f \label{eq: acce}
\end{equation}

\subsection{Immersed boundary method}
In the IBM, it is assumed that the viscous fluid fills both the inside and outside of the immersed boundary. To enforce the no-slip boundary condition, localized body forces are applied to the Eulerian lattice points near the fluid-solid interface. The specific methodology for determining these body forces varies among different IBM implementations~\cite{jiang2019boundary}. In this work, we adopt the multi-direct-forcing scheme. This method is essentially an iterative extension of the standard direct-forcing scheme, designed to iteratively reduce the velocity slip at the boundary for enhanced accuracy. 
Let $\mathbf{X}_k$ and $\mathbf{U}_k$ denote the coordinates and the prescribed velocities of the $N$ Lagrangian markers representing the moving boundary, respectively. Since the Lagrangian points $\mathbf{X}_k$ generally do not coincide with the Eulerian grid points $\mathbf{x}$, interpolations between the two systems are required. 
Given the distribution functions $f_i$ at time $t$, an intermediate unforced velocity field $\mathbf{u}^*$ is first obtained using Eq.~(\ref{eq: cal_velocity}) without the IBM forcing term. Subsequently, the momentum exchange between the boundary and the fluid is realized by the following four-step procedure, which is iterated multiple times within a single time step in the multi-direct-forcing scheme \cite{suzuki_effect_2011}:
\begin{itemize}
    \item interpolation of the velocity from the Eulerian grid $\mathbf{x}$ to the position of Lagrangian markers $\mathbf{X}_k(t)$:
\begin{equation}
    \mathbf{u}^*_k(\mathbf{X}_k, t) = \sum_{\mathbf{x}}\mathbf{u}^*(\mathbf{x},t)W(\mathbf{x}-\mathbf{X}_k)(\Delta x)^D \label{eq: ibm_interpolate}
\end{equation}
    \item computation of the body force term $\mathbf{F}_k$ for each Lagrangian marker $\mathbf{X}_k(t)$ according to the velocity of the immersed boundary, $\mathbf{U}_k$:
\begin{equation}
    \mathbf{F}_k(\mathbf{X}_k, t) = \rho_f(\mathbf{U}_k(\mathbf{X}_k, t)-\mathbf{u}_k^*(\mathbf{X}_k, t))/\Delta t
\end{equation}
    \item Spread, i.e. extrapolate, the force $\mathbf{F}_k$ from the Lagrangian markers $\mathbf{X}_k$ to the Eulerian grid $\mathbf{x}$:
\begin{equation}
    \mathbf{F}(\mathbf{x},t) = \sum_{k=1}^{N}\mathbf{F}_k(\mathbf{X}_k,t)W(\mathbf{x}-\mathbf{X}_k)(\Delta x)^D
\end{equation}
    \item Correct the velocity at the Eulerian grid point:
\begin{equation}
    \mathbf{u}(\mathbf{x},t) = \mathbf{u}^*(\mathbf{x},t) + \mathbf{F}(\mathbf{x},t)\Delta t/\rho_f
\end{equation}    
\end{itemize}
Where the function $W$ is a weighting function proposed by Peskin \cite{Peskin2002}, and $D$ is the dimensionality. The weighting function $W$ is given by:
\begin{equation}
W(x,y,z) = \frac{1}{\Delta x}\phi(\frac{x}{\Delta x})\cdot \frac{1}{\Delta x}\phi(\frac{y}{\Delta x})\cdot \frac{1}{\Delta x}\phi(\frac{z}{\Delta x})
\end{equation}
Here, the 3-point discrete delta function $\phi(x)$ is employed, which is given by:
\begin{equation}
\phi(x) = 
\begin{cases}
\frac{1}{3}(1+\sqrt{1-3x^2}) & 0\leq|x|\leq 0.5 \\
\frac{1}{6}(5-3|x|-\sqrt{-2+6|x|-3x^2}) & 0.5\leq|x|\leq 1.5 \\
0 & 1.5\leq|x| 
\end{cases}
\end{equation}

\subsection{Dynamics of the Dispersed Phases}
\label{sec:suspended_phases}
The multiphase system considered in this work consists of two distinct dispersed phases: finite-size bodies fully resolved by the IBM, and small particles modeled using the point-particle model.

\begin{figure}[ht]
\centering
\includegraphics[width=1.0\linewidth]{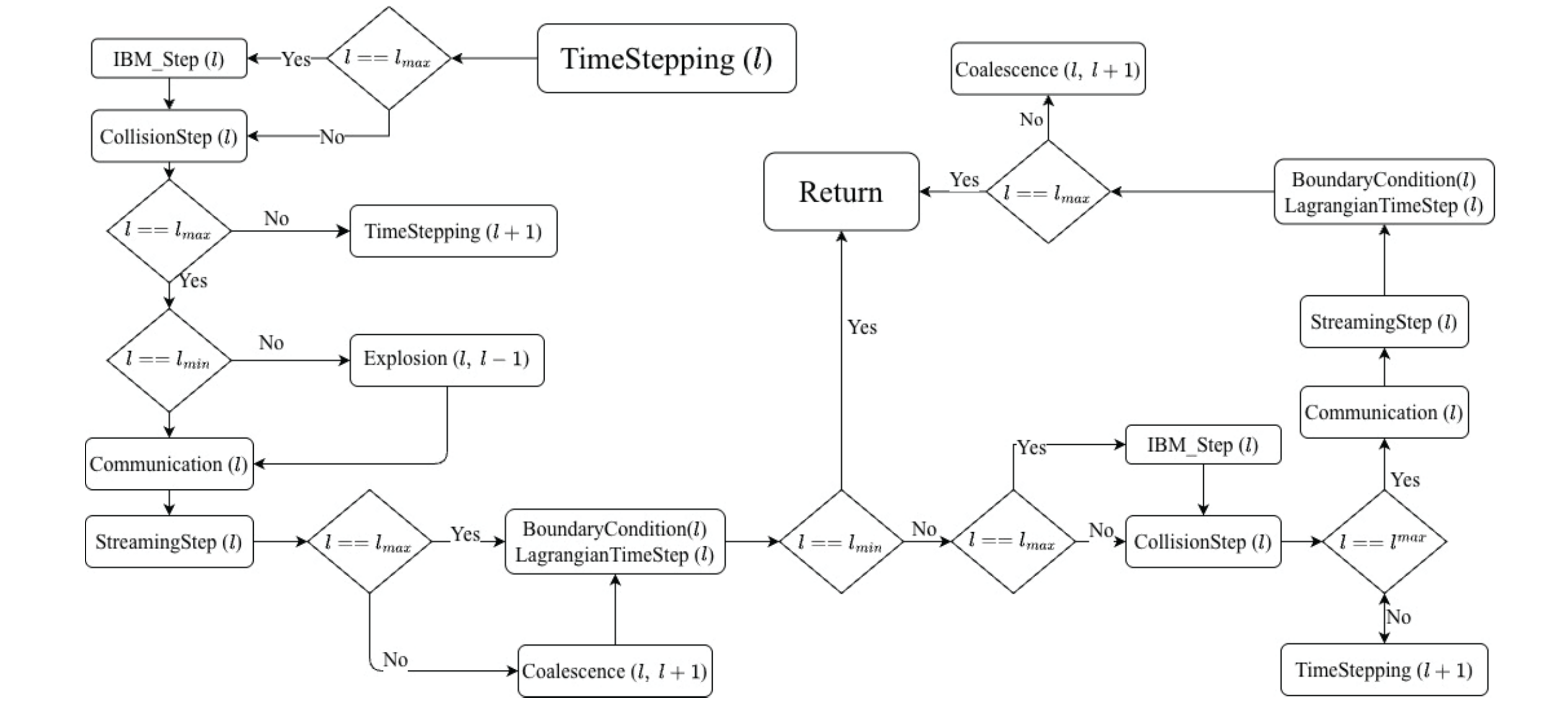}
\caption{Flowchart of the recursive time-stepping algorithm for the coupled LBM-IBM solver on adaptive grids. The diagram illustrates how a coarse time step $\Delta t_l$ is decomposed into two finer sub-steps $\Delta t_{l+1}$.
The IBM coupling operations are exclusively executed at the finest grid level ($l_{max}$).}
\label{fig:time_stepping}
\end{figure}

\subsubsection{Fully Resolved Finite-size Bodies}
In the present framework, the finite-size dispersed entities are modeled as rigid bodies.
The translational and rotational dynamics of the finite-size bodies are governed by the Newton--Euler equations. Let $\mathbf{x}_b$, $\mathbf{v}_b$, and $\mathbf{\Omega}_b$ denote the centroid position, translational velocity, and angular velocity, respectively, of a body with mass $m_b$, volume $V_b$, and moment of inertia tensor $\mathbf{I}_b$. 
The equations of motion are:
\begin{eqnarray}
m_b \frac{\mathrm{d} \mathbf{v}_b}{\mathrm{d}t} &=& \mathbf{F}_{hyd} + (m_b - \rho_f V_b)\mathbf{g}, \label{eq:Newton-Euler1}\\
\frac{\mathrm{d} (\mathbf{I}_b  \mathbf{\Omega}_b)}{\mathrm{d}t} &=& \mathbf{T}_{hyd}, \label{eq:Newton-Euler2}  
\end{eqnarray}
where $\mathbf{g}=-g\mathbf{e}_z$ is the gravitational acceleration vector. The hydrodynamic force $\mathbf{F}_{hyd}$ and torque $\mathbf{T}_{hyd}$ acting on the particle surface are computed directly from the immersed boundary forcing. In IBM, the force exerted by the fluid on the structure is the reaction to the force exerted by the structure on the fluid, which is given by the summation over all Lagrangian markers, $\mathbf{F}^{ibm} = -\sum_{k=1}^N\mathbf{F}_k\Delta V_k $ ($\Delta V_k$ is the volume element defined as $S/N\times \Delta x$ where $S$ is the total surface area of the body). 
However, this term is in general not equal to $\mathbf{F}_{hyd}$, because a part of the forcing is used to drive the internal fluid inside the finite-size body \cite{Uhlmann2005, feng2009robust}. We note that this effect is always significant, and must be taken into account for an accurate dynamics description, independently of the mass density of the particle. 
These internal compensation forces and torques can be obtained by the time rate variation of linear momentum $\mathbf{P}$ and angular momentum $\mathbf{L}$ of the fluid internally located in the particle.
Thus, we have 
\begin{eqnarray}
\mathbf{F}_{hyd}=-\sum_{k=1}^N\mathbf{F}_k\Delta V_k + \frac{{\rm d}\mathbf{P}(t)}{{\rm d}t} \\
\mathbf{T}_{hyd}=-\sum_{k=1}^N(\mathbf{x}-\mathbf{x}_b)\times\mathbf{F}_k\Delta V_k + \frac{{\rm d}\mathbf{L}(t)}{{\rm d}t}
\end{eqnarray}
An additional set of Lagrangian points $\mathbf{X}_{in}$ are located inside the body and move with it. They are used to calculate the linear and angular momenta of the internal fluid as follows: 
\begin{eqnarray}
\mathbf{P}(t)=\rho_f\int_{x\in\Gamma} \mathbf{u}(\mathbf{x},t){\rm d}\mathbf{x}&\simeq&\sum_{all\,\mathbf{X}_{in}}\mathbf{u}(\mathbf{X}_{in},t)\Delta V_{in}\\
\mathbf{L}(t) = \rho_f\int_{x\in\Gamma} (\mathbf{x}-\mathbf{x}_b)\times\mathbf{u}(\mathbf{x},t){\rm d}\mathbf{x} &\simeq& \sum_{all\,\mathbf{X}_{in}}[\mathbf{X}_{in}(t)-\mathbf{x}_b(t)]\times\mathbf{u}(\mathbf{X}_{in},t)\Delta V_{in}.
\end{eqnarray}
where $\Gamma$ is the closed domain inside the finite-size body. Here, we set the initial internal Lagrangian markers $\mathbf{X}_{in}(0)$ with the width of $\Delta x_{l=max}$, therefore, we take $\Delta {V}_{in} =(\Delta x_{l=max})^D$.

\begin{figure}[ht]
\centerline{\includegraphics[width=1.0\linewidth]{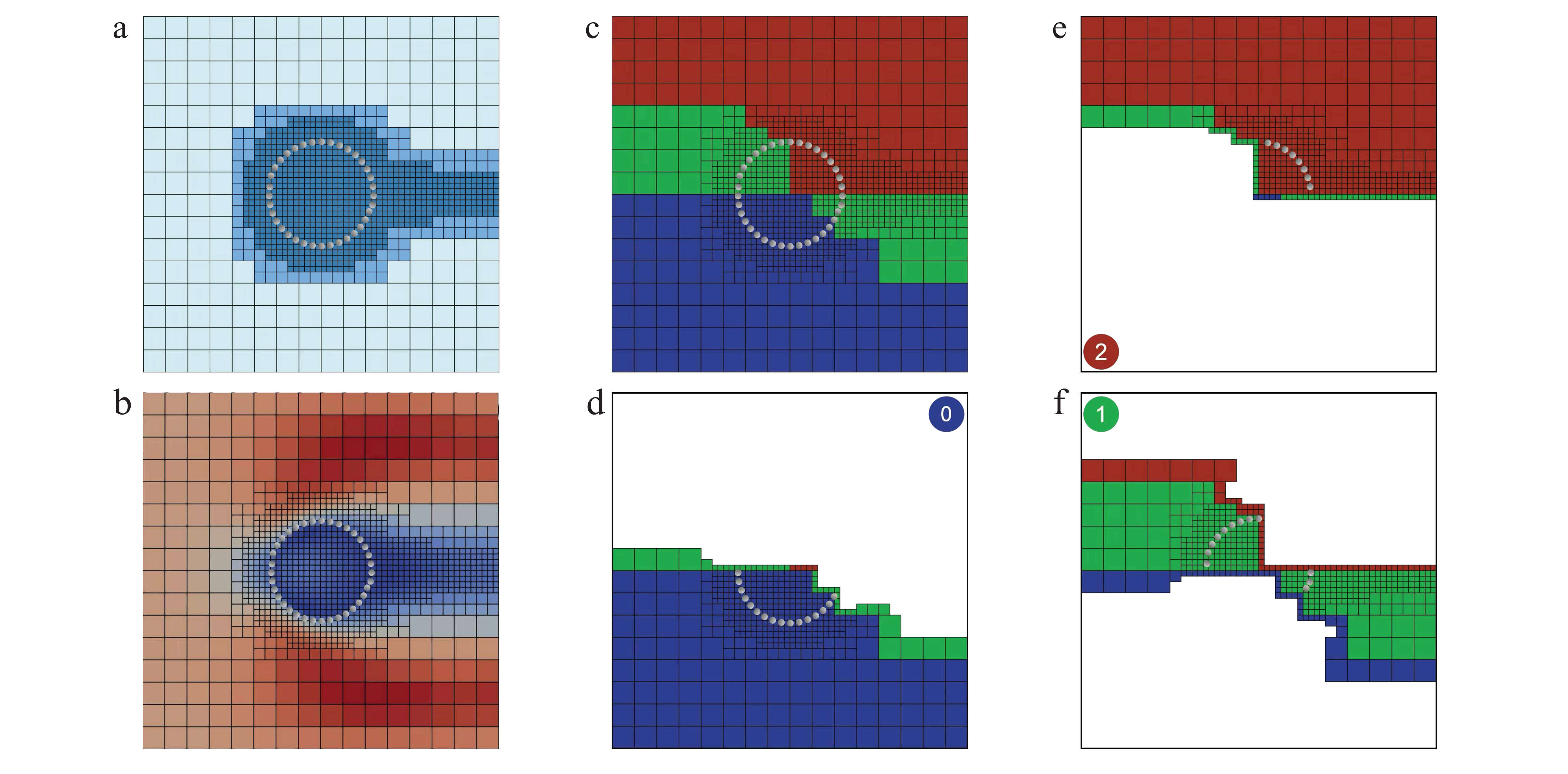}}
\caption{Illustration of the parallel adaptive grid configuration for the LBM-IBM simulation of flow past a cylinder. (a) Grid-level distribution with three refinement levels. High resolution is concentrated near the cylinder surface (represented by Lagrangian markers $\mathbf{X}_k$, shown as dots) and in the wake region. (b) Snapshot of the velocity field; note the grid is coarsened for visualization. (c) Domain decomposition among three MPI processes (Rank 0: Blue, Rank 1: Green, Rank 2: Red). (d–f) The local grid forests belonging to each MPI process. A layer of ghost cells from neighboring processes is explicitly maintained to enable the exchange of distribution functions at partition boundaries.}
\label{fig:meshshow}
\end{figure}

\subsubsection{Point-like Particles}
In contrast to the finite-size particles, whose size is much larger than the typical flow scale, particles with sizes comparable to or smaller than $\eta$, can be modeled as point-like particles. 
It should be noted that extending
the current framework to two-way momentum coupling is not trivial on adaptive grids. If the particle size exceeds the local cell size, proper filtering operators akin to the discrete delta
function \cite{capecelatro2013euler} are required to avoid numerical singularities.
In this study, a truncated Maxey-Riley-Gatignol (MRG) equation is employed. Based on the literature, hydrodynamic drag and gravity predominantly dictate the macroscopic collision process \cite{huang_effect_2012}. However, we acknowledge that as small particles approach the bubble surface, additional terms such as lift forces and sub-grid lubrication forces may alter the near-wall trajectories. Incorporating these micro-physical forces, alongside the Basset history term, represents an important direction for future extensions.
The motion of a particle with density $\rho_p$ and radius $r_p=d_p/2$ is  governed by the balance of inertia, hydrodynamic drag, and net gravity (buoyancy):
\begin{equation}
m_p \frac{\mathrm{d}\mathbf{v}_p}{\mathrm{d}t} = 6\pi\mu r_p f_p (\mathbf{u}(\mathbf{x}_p) - \mathbf{v}_p) + (m_p - m_f)\mathbf{g},
\label{eq:PointParticle}
\end{equation}
where $\mathbf{v_p}$ is the particle velocity, $\mathbf{u}(\mathbf{x}_p)$ is the fluid velocity at the particle location, and $m_p=\frac{4}{3}\pi r_p^3\rho_p$ is the particle mass.
The drag coefficient $f_p=1+0.15Re_p^{0.687}$ is the correction factor that accounts for nonlinear drag and is a function of particle Reynolds number ($Re_p=2r_p|\mathbf{u}-\mathbf{v}_p|/\nu$). 
The particle equations are integrated using a second-order Adams--Bashforth (AB2) scheme, with the fluid velocity $\mathbf{u}$ obtained via trilinear interpolation from the Eulerian grid.  
Crucially, the time integration is tightly coupled with the adaptive fluid solver: the particle time step $\Delta t_p$ is dynamically adjusted to match the local fluid time step $\Delta t_l$ corresponding to the refinement level $l$ of the host cell where the particle currently resides. This ensures strict temporal synchronization between the dispersed phase and the multi-level carrier flow.
For the current one-way coupling, the fluid velocity at the exact location of the small point particles is obtained via trilinear interpolation from the surrounding Eulerian grid nodes.
A specific numerical challenge arises from the multi-step nature of the AB2 scheme, which relies on the right-hand side evaluation (i.e., the particle acceleration) from the previous time step. Because the adaptive framework employs level-dependent lattice units, a particle migrating across a refinement boundary carries the acceleration data scaled in the units of its previous host cell. To preserve physical and mathematical consistency, this previous-step acceleration must be explicitly rescaled (as per the acceleration scaling laws discussed in Section~\ref{sec:LBM}) to the local lattice units of the new refinement level prior to executing the AB2 time integration.

\subsection{Recursive Time-Integration Scheme on Adaptive Grids}
\label{sec:time_marching}
The temporal evolution of the coupled system is managed by a recursive time-stepping algorithm, following the strategy implemented in the \texttt{waLBerla} framework~\cite{bauer2021walberla}. This scheme orchestrates the sub-cycling of the fluid solver, the force coupling of the IBM, and the particle updates across different refinement levels.

As illustrated in Fig.~\ref{fig:time_stepping}, the algorithm performs a depth-first traversal of the grid hierarchy. For a simulation with a maximum refinement level $l_{max}$ and a base level $l_{min}$, a  single time step on a coarse level $l$ involves two recursive sub-steps on the finer level $l+1$. 
This ensures that the acoustic scaling constraint ($\Delta t_l=2\Delta t_{l+1}$) is satisfied.

A critical design choice in our framework is the treatment of the finite-size body. The immersed boundary calculations (velocity interpolation and force spreading) are strictly restricted to the finest grid level $l_{max}$. 
Restricting the IBM operations to the finest grid level not only facilitates the consistent application of the discrete delta function but is also physically motivated, as the highest shear gradients are localized at the particle interface.
Consequently, the IBM coupling routines are only invoked when the recursive cycle reaches the leaf nodes at the maximum depth of the tree.
In contrast, the point-particle integration (as described in Section~\ref{sec:suspended_phases}) is performed at the local level where the particle resides, ensuring synchronization with the local fluid time step.

The synchronization between levels, specifically the exchange of distribution functions via the virtual cell layers and the coarse-fine interpolation, is performed immediately after the collision-streaming cycles of the respective levels, ensuring flux conservation across the adaptive mesh interfaces.

\section{Parallelization Approach and Implementation}
\label{sec:parallel}
The present code structure employs a fully parallelized algorithm designed to efficiently handle large-scale simulations involving complex fluid–structure interactions. 
Rather than modifying existing CFD codes
(e.g. Basilisk or waLBerla), we developed a dedicated in-house solver based on the \texttt{p4est} library. This provides the direct data control needed to efficiently integrate our adaptation algorithms and auxiliary grids for multi-scale particle simulations.
The implementation focuses on ensuring scalability, minimizing communication overhead, and maintaining numerical consistency across distributed processes. This section outlines the core components of the parallel strategy, structured into three parts: (i) Grid management, which details the domain decomposition, dynamic adaptation, and data communication on the Eulerian mesh; (ii) Lagrangian data handling, focusing on the parallel storage and transfer of finite-size particles and Lagrangian markers in IBM; and (iii) The auxiliary grid forest constructed for the IBM, which facilitates the coupling between Eulerian and Lagrangian representations, and the computational efficiency.

\subsection{Grid Management and Domain Decomposition}
The use of adaptive tree-based grids significantly reduces the computational cost of simulating finite-size particle-laden flows. By restricting fine grid resolution to the immediate vicinity of particle boundary layers where high gradients exist, we could avoid over-resolving the background flow field, i.e., turbulence. The cell-centered LBM formulation adopted here is naturally compatible with tree-based data structures, allowing for efficient encoding of the grid hierarchy~\cite{mehl2019adaptive}.
In this work, the computational domain is managed using the \texttt{p4est} library~\cite{burstedde2011p4est}. This library implements a dynamic forest of octrees (in 3D) or quadtrees (in 2D), enabling the domain to be viewed as a collection of trees that can be individually refined or coarsened. This structure supports massively parallel execution by leveraging SFCs, specifically the Z-curve (Morton order), to map the multi-dimensional tree structure onto a linear sequence of processes.
The grid management, as illustrated in Fig.~\ref{fig:meshshow}, involves three fundamental operations: \textit{refinement/coarsening}, \textit{balancing}, and \textit{partitioning}.
\begin{enumerate}
\item \textbf{Refinement/Coarsening:} The grid resolution is dynamically adjusted based on user-defined criteria. As shown in Fig.~\ref{fig:meshshow}(a-b) for a flow past a cylinder, cells are tagged for refinement in regions of high velocity gradients or proximity to the immersed boundary.
\item \textbf{Balancing:} A 2:1 balance constraint is enforced for the algorithm of LBM in adaptive grids. This ensures that the size ratio between any two adjacent cells does not exceed 2, which is critical for the validity of the virtual cell interpolation schemes described in Section~\ref{sec:LBM}.
\item \textbf{Partitioning:} The computational load is distributed among MPI processes by cutting the linear SFC segment. Because SFCs preserve spatial locality, this method ensures that cells on the same process are geometrically close, thereby minimizing inter-process communication volume. Fig.~\ref{fig:meshshow}(c-f) illustrates this domain decomposition for three MPI ranks.
\end{enumerate}

To facilitate the LBM streaming step across process boundaries, a layer of \textit{ghost cells} is maintained (see Fig.~\ref{fig:meshshow}(d-f)). These ghost cells store copies of the particle distribution functions $f_i$ from neighboring ranks. During each time step, a communication routine synchronizes these ghost layers, ensuring that the local LBM update remains consistent with the global flow field. 
The domain partitioning used is not purely geometric; it employs a weighted load balancing strategy, assigning weights proportional to the number of sub-cycles (weight $\propto 2^{\rm level}$). While this effectively balances the computational collision-stream workload across MPI ranks, it can lead to less compact partition geometries, increasing the communication-to-computation ratio and introducing a practical trade-off with MPI communication overhead.
It should be noted that the grid management operations described in this section are standard implementations based on the \texttt{p4est} library. They are outlined here strictly to establish the baseline adaptive topology required to introduce the novel coupling algorithms in the following sections.

\begin{algorithm}[ht]
\caption{search\_host\_cell(points): Algorithm for locating the cell containing a Lagrangian point, where $MAXLEVEL=30$ refers to the maximum finest level for representing cells in $p4est$}\label{alg:lagrangian_searching}
{\small
\begin{algorithmic}[1]
\For{$point\,\mathbf{in}\, points$}
\State $isFound\gets false$
\For{$level\,\mathbf{in}\, [l^{min},l^{max}]$}
\State $test\_cell = get\_cell\_at(point,\, level)$ \Comment Create a test cell based on the $point$ location and  $level$
    
\For{$tree\,\mathbf{in}\, trees$}
    \State $idx \gets {\rm binary\_search}(tree.{\rm quadrants}, test\_cell)$ \Comment Perform a binary search on the given $tree$ based on Z-order
    \If{$idx \geq 0$}
    \State $point.{\rm host\_cell\_id} \gets idx$
    \State $point.{\rm which\_tree} \gets tree$
    \State $isFound\gets true$
    \State $break$
    \EndIf
\EndFor
    \If{$isFound$}
    \State $break$
    \EndIf
\EndFor
\EndFor
\Function{$get\_cell\_at(point, level)$}
    \State $test\_cell.{\rm level} \gets level$
    \State $test\_cell.{\rm x} \gets 
    \bigl\lfloor point.{\rm x} \cdot 2^{\text{level}} \bigr\rfloor 
    \cdot 2^{\text{MAXLEVEL} - \text{level}}$
    \State $test\_cell.{\rm y} \gets 
    \bigl\lfloor point.{\rm y} \cdot 2^{\text{level}} \bigr\rfloor 
    \cdot 2^{\text{MAXLEVEL} - \text{level}}$
    \State $test\_cell.{\rm z} \gets 
    \bigl\lfloor point.{\rm z} \cdot 2^{\text{level}} \bigr\rfloor 
    \cdot 2^{\text{MAXLEVEL} - \text{level}}$
    \State \textbf{return} test\_cell
\EndFunction

\end{algorithmic}
}
\end{algorithm}

\subsection{Lagrangian Data Management}
\label{sec:lagrangian_method}
In our simulation framework, the Lagrangian phase encompasses four distinct categories of physical and numerical entities: (i) the point-like particles; (ii) the representations for finite-size bodies; (iii) Lagrangian markers $\mathbf{X}_k$ in IBM, defining the immersed boundaries; and (iv) internal Lagrangian points $\mathbf{X}_{in}$, used for the calculation of the linear and angular momentum of the IBM body.
 
Given the potentially massive number of Lagrangian points ($N_p$), a distributed memory approach is essential. Each MPI process exclusively stores and updates the particles located within its local spatial subdomain to ensure memory scalability. A fundamental operation in this coupled framework is the host cell identification: for every time step, each Lagrangian point must be associated with the Eulerian grid cell (host cell) in which it is located in. This association is a prerequisite for interpolating flow quantities (e.g., fluid velocity) to the particle and for spreading particle forces back to the fluid. 
On uniform structured grids, mapping a particle coordinate  to a cell index is a trivial $O(1)$ operation. However, on an unstructured or adaptive octree grid, the problem is non-trivial. A naive linear search, which checks every local leaf cell to see if it contains the particle, results in a computational complexity of $O(N_p N_{cell})$ ($N_{cell}$ is the number of grids). As the number of cells and particles grows, this approach becomes prohibitively expensive. 

Following the standard algebraic search paradigm in SFC-based frameworks \cite{burstedde2011p4est, isaac2015recursive, burstedde2020parallel}, the core idea is that the leaf cells are stored in a linearized array sorted by their Morton (Z-order) index. Therefore, instead of a geometric search, we perform an algebraic binary search: 
(i) Compute the leaf cell for the particle position at a given level; (ii) Perform a binary search over the sorted array of local leaf cells to find the cell that encompasses the cell; (iii) adjust the refinement level and repeat until the host cell is found. This strategy reduces the complexity from $O(N_p N_{cell})$ to $O(N_p\log N_{cell})$, ensuring scalability even for dense particle suspensions. The procedure is detailed in Algorithm~\ref{alg:lagrangian_searching}.

\begin{figure}[ht]
\centerline{\includegraphics[width=1.0\linewidth]{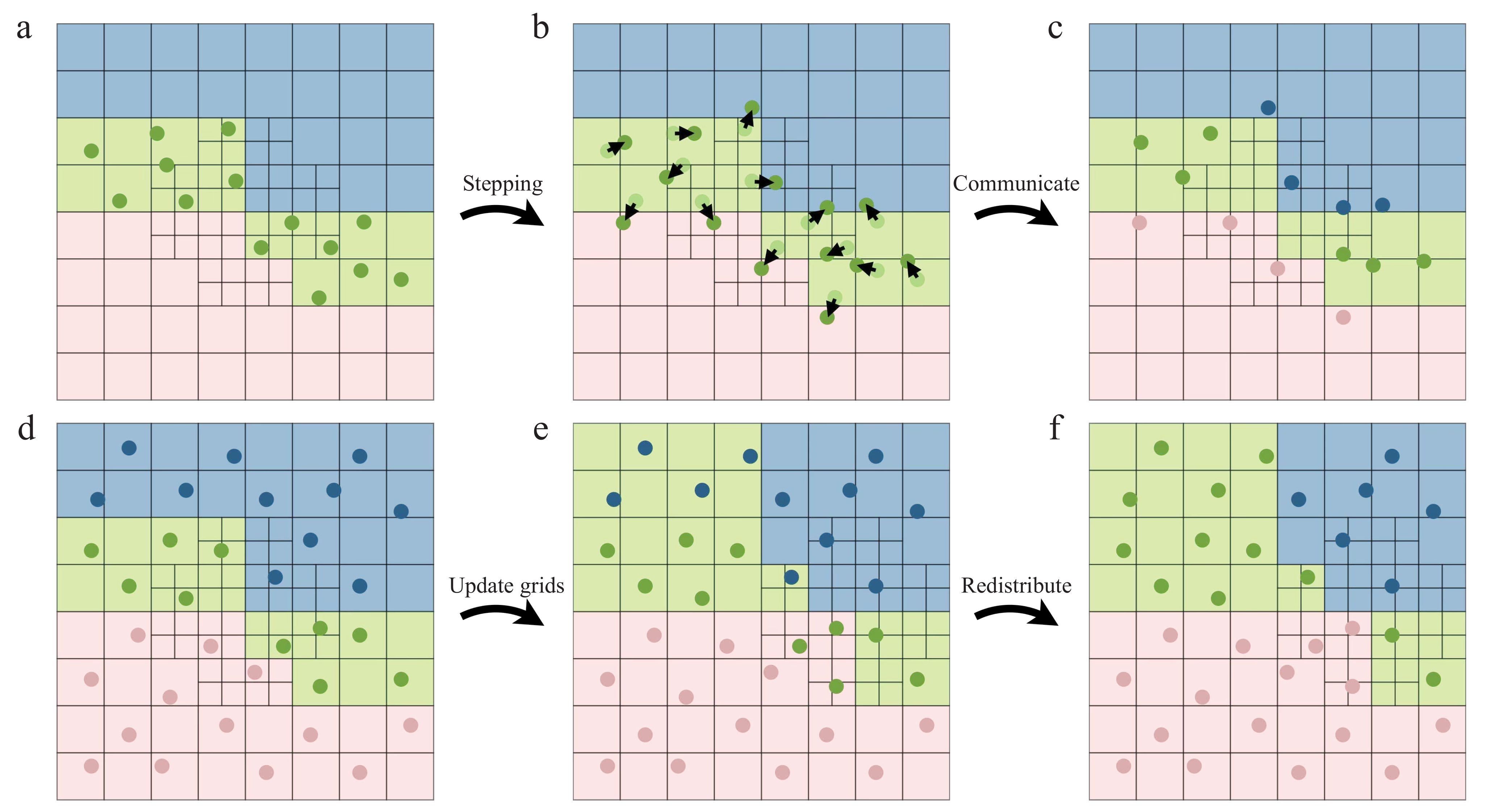}}
  \caption{Illustration of two types of parallel processes of Lagrangian points. Panels (a–c) depict the communication of Lagrangian points between neighboring processes. Panels (d–f) show the redistribution of Lagrangian points across different MPI processes after the mesh update. The colors of the grid cells and points indicate the MPI ranks to which they belong, with pink, green, and blue representing ranks 1, 2, and 3, respectively.  }
  \label{fig:lagrange_parallel}
\end{figure}

\begin{figure} [h]
\centering
\includegraphics[width=0.4\linewidth]{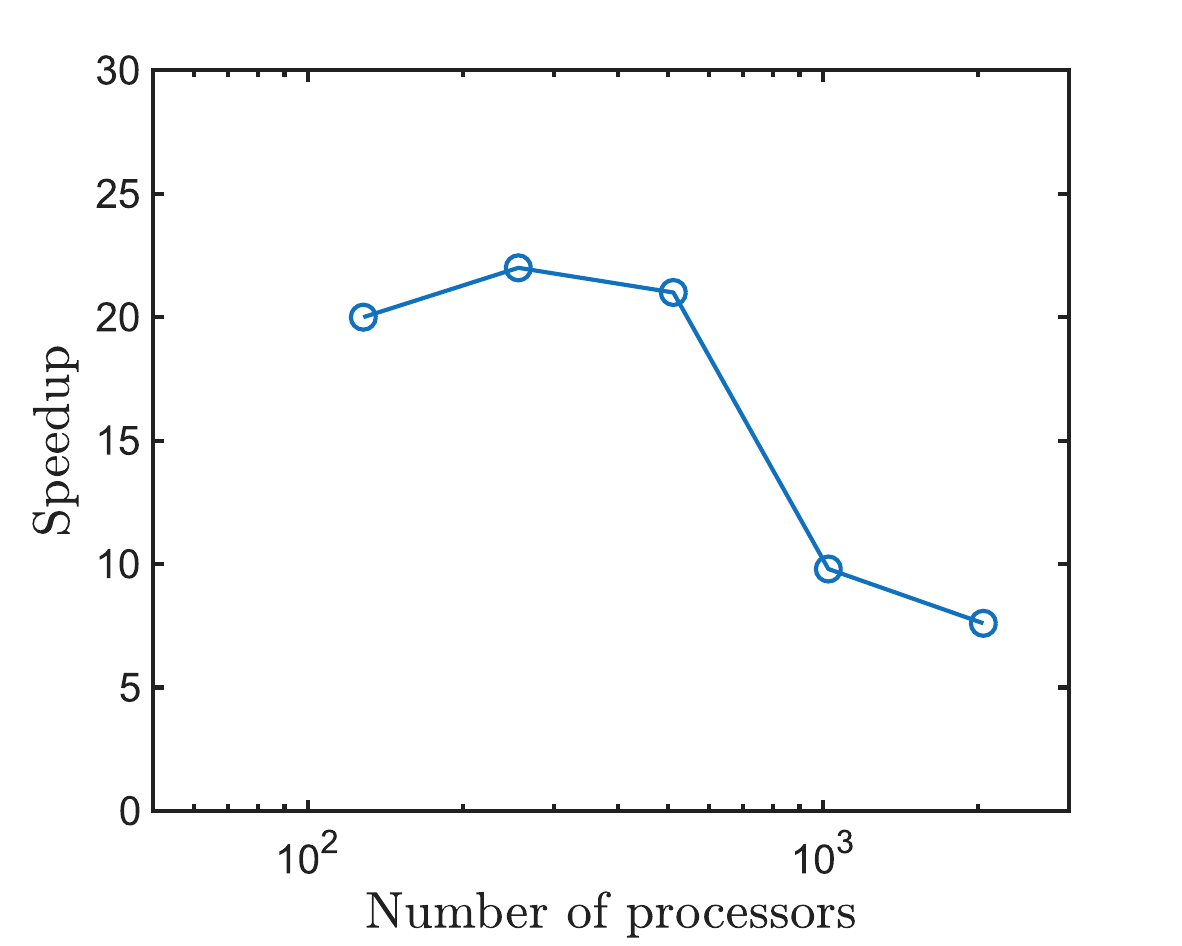}
\caption{The speedup of the binary searching algorithm for the Lagrangian data communication step compared with the communication step using plain linear searching.}
\label{fig:speedup}
\end{figure}

In a dynamic adaptive environment, particle data must be migrated between processors to maintain data locality. We distinguish between two scenarios requiring parallel communication, as illustrated in Fig.~\ref{fig:lagrange_parallel}:
\begin{enumerate}
\item \textbf{Post-Advection Communication (Particle Crossing):}
During a time step, particles update their positions based on the local velocity. A particle may cross the boundary of its current partition. Since the time step is limited by the convergence condition, particles typically move only a short distance (less than one grid cell). Therefore, particles leaving the local domain will invariably land in the \textit{ghost layer} region (see Fig.~\ref{fig:meshshow}(d-f)). The communication strategy involves (see Fig.~\ref{fig:lagrange_parallel}(a-c)):
\begin{itemize}
\item Identify particles located in ghost cells.
\item Pack these particles into buffers destined for the corresponding neighbor rank (owner of the ghost cell).
\item Exchange data via MPI non-blocking point-to-point communication.
\end{itemize}
\item \textbf{Post-Refinement Redistribution (Grid Changing):} 
When the mesh is refined, coarsened, or repartitioned for load balancing, the domain boundaries shift. A particle that was stationary may suddenly find itself belonging to a different MPI process because its host cell has been reassigned (see Fig.~\ref{fig:lagrange_parallel}(d-f)). Unlike the advection case, the new owner might not be a direct spatial neighbor. To handle this:
\begin{itemize}
    \item For every particle, we compute its Morton code and determine the expected owner rank by comparing the code against the global SFC partition range of each process (which is available on all ranks).
    \item Particles that no longer belong to the local rank are collected and migrated to their new owners.
\end{itemize}
\end{enumerate}
Details of the procedures are illustrated in Algorithm \ref{alg:lagrangian_commu}. Both procedures ensure that after every sub-step of the simulation, the Lagrangian particles are strictly resident on the process that owns their Eulerian host cell, preserving the efficiency of the coupled solver.
To quantitatively evaluate the impact of the search algorithm on the parallel performance, we conduct a benchmark simulation featuring a simple shear flow in a domain discretized by $64^3$ grid cells, uniformly seeded with $N_p = 10^7$ Lagrangian point particles. Figure~\ref{fig:speedup} presents the computational speedup achieved specifically during the Lagrangian data communication step, compared against a baseline utilizing a naive linear search. The results demonstrate a tremendous acceleration in the overall data migration process. 
It is worth noting that the speedup declines as the number of processors increases. This is primarily because the local grid size per process shrinks, which reduces the computational burden of the naive linear search, thereby narrowing the efficiency gap between the two algorithms. Furthermore, at higher processor counts, the fixed overhead of MPI data transfer accounts for an increasingly larger fraction of the total communication step, diluting the relative computational speedup gained solely from the optimized host-cell identification.
This highlights that optimizing the host-cell identification effectively eliminates the primary bottleneck in particle communication, directly reflecting the algorithmic improvement from $O(N_p N_{cell})$ to $O(N_p \log N_{cell})$, and thereby confirms the efficiency of the implemented search strategy for managing and communicating dense particle suspensions in parallel environments.

\begin{algorithm} [h]
\caption{Parallel migration and communication of Lagrangian points among processes. The unified procedure handles both particle advection and adaptation through three main phases. (I) Pre-search: identify outbound points by either detecting crossings into ghost cells (for post-advection) or querying the global SFC partition map (for post-refinement redistribution). (II) Communication: exchange point data with the determined target processes. (III) Post-search: process the incoming points and determine the corresponding host cells in which they reside.}\label{alg:lagrangian_commu}
{\small
\begin{algorithmic}[1]
\State \textbf{(I) Pre-search: Identify outbound points and target processes}
\State Initialize $SendBuffer$ and $RecvBuffer$
\For {$point$ $\mathbf{in}$ $points$}
    \State search\_host\_cell($point$) \Comment{Locate the local host cell of the point using Algorithm~\ref{alg:lagrangian_searching}}
    \If{$point.{\rm host\_cell\_id}>0$}    \Comment{A valid local host cell is found}
        \State continue  \Comment{Keep the $point$ in the local array}
    \EndIf
    \State $ownerRank \gets {\rm find\_owner}(point)$  \Comment{Identify the owner rank of the $point$ (in ghost layer or on a remote process)}
    \State $SendBuffer[ownerRank].{\rm push\_back}(point)$ \Comment{Append the point data to the corresponding send buffer}
    \State $points.{\rm erase}(point)$   \Comment{Remove this $point$ from the local $points$ array}
\EndFor

\vspace{0.2cm}
\State \textbf{(II) Communication: Exchange data}
\State Exchange buffer sizes with corresponding target processes
\State $\text{MPI\_Isend} / \text{MPI\_Irecv} (SendBuffer \leftrightarrow RecvBuffer)$ \Comment{Perform non-blocking send/receive of buffers}
\State $\text{MPI\_Waitall}()$ \Comment{Wait until all communication requests are completed}

\vspace{0.2cm}
\State \textbf{(III) Post-search: Process incoming points}
\For{$point$  $\mathbf{in}$ $RecvBuffer$}
    \State search\_host\_cell($point$) \Comment{Determine the exact local host cell of the incoming point using Algorithm~\ref{alg:lagrangian_searching}}
    \State $points.{\rm push\_back}(point)$ \Comment{Add the $point$ to the local $points$ array}
\EndFor

\end{algorithmic}
}% end font small
\end{algorithm}

\subsection{Auxiliary Grid for Efficient Lagrangian Tracking}
\label{sec:aux_grid}
\begin{figure}[ht]
\centerline{\includegraphics[width=1.0\linewidth]{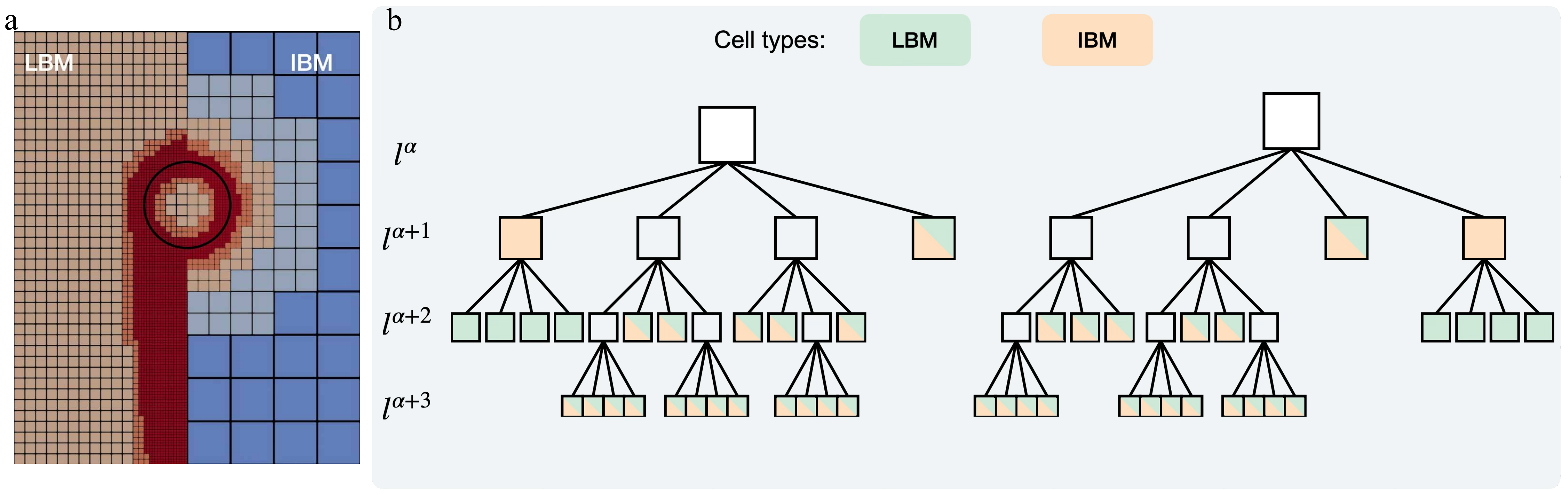}}
  \caption{Comparison of the primary and auxiliary grid structures. (a) Representation of the grid cells: The left side shows the LBM fluid grid, which is refined in the wake to capture flow dynamics; the right side illustrates the auxiliary IBM grid, which is refined \textit{only} near the particle interface (black line). (b) Hierarchical branching diagram of the corresponding octree data structures. The second-level nodes in the tree diagram correspond to the base-level cells of the LBM tree shown in (a). The active leaf cells are highlighted in color (green for the primary LBM tree and pink for the auxiliary IBM tree), whereas the intermediate, non-leaf cells are left uncolored. The auxiliary IBM tree exhibits a significantly reduced active cell count compared to the primary LBM tree, which facilitates much faster tree traversal and particle localization.}
  \label{fig:ibm_octree}
\end{figure}

In standard LBM-IBM implementations, the Lagrangian markers typically interact directly with the primary fluid grid. However, in turbulent flow simulations, the fluid grid often requires extensive refinement in wake regions and shear layers far from the particles themselves. This results in a massive number of Eulerian leaf cells ($N_{fluid}$), which significantly increases the computational cost of the host cell search algorithm described in Section~\ref{sec:lagrangian_method}, as the search complexity scales with $log(N_{fluid})$.
To overcome this bottleneck, we introduce a dual-mesh strategy. In addition to the primary adaptive mesh for the LBM solver, an auxiliary grid forest is constructed specifically for the immersed boundary method, leveraging the \texttt{p4est\_build} infrastructure. As illustrated in Fig.~\ref{fig:ibm_octree}, this auxiliary forest is structurally distinct from the fluid mesh:
(i) Sparsity: The auxiliary grid is designed to be globally sparse. It remains at the coarsest possible level throughout the domain, except in the immediate vicinity of the Lagrangian markers in IBM.
(ii) Targeted Refinement: Refinement in the auxiliary grid is driven strictly by the proximity to the IBM surface, independent of flow features like vorticity or velocity gradients.

This strategy serves two critical purposes.
First and foremost, it drastically accelerates the geometric search for Lagrangian points. Since the auxiliary forest ignores the complex refinement requirements of the background turbulence, its total cell count is orders of magnitude smaller than that of the fluid grid. Consequently, the host cell identification becomes significantly faster. 
Second, it simplifies the implementation of IBM-specific algorithms, enabling the Lagrangian–Eulerian coupling to be handled independently from the fluid solver’s adaptive mesh hierarchy. This separation provides flexibility in performing interpolation and spreading operations without interfering with the data structures of the LBM solver.

\begin{figure}[ht]
 \centerline{\includegraphics[width=0.8\linewidth]{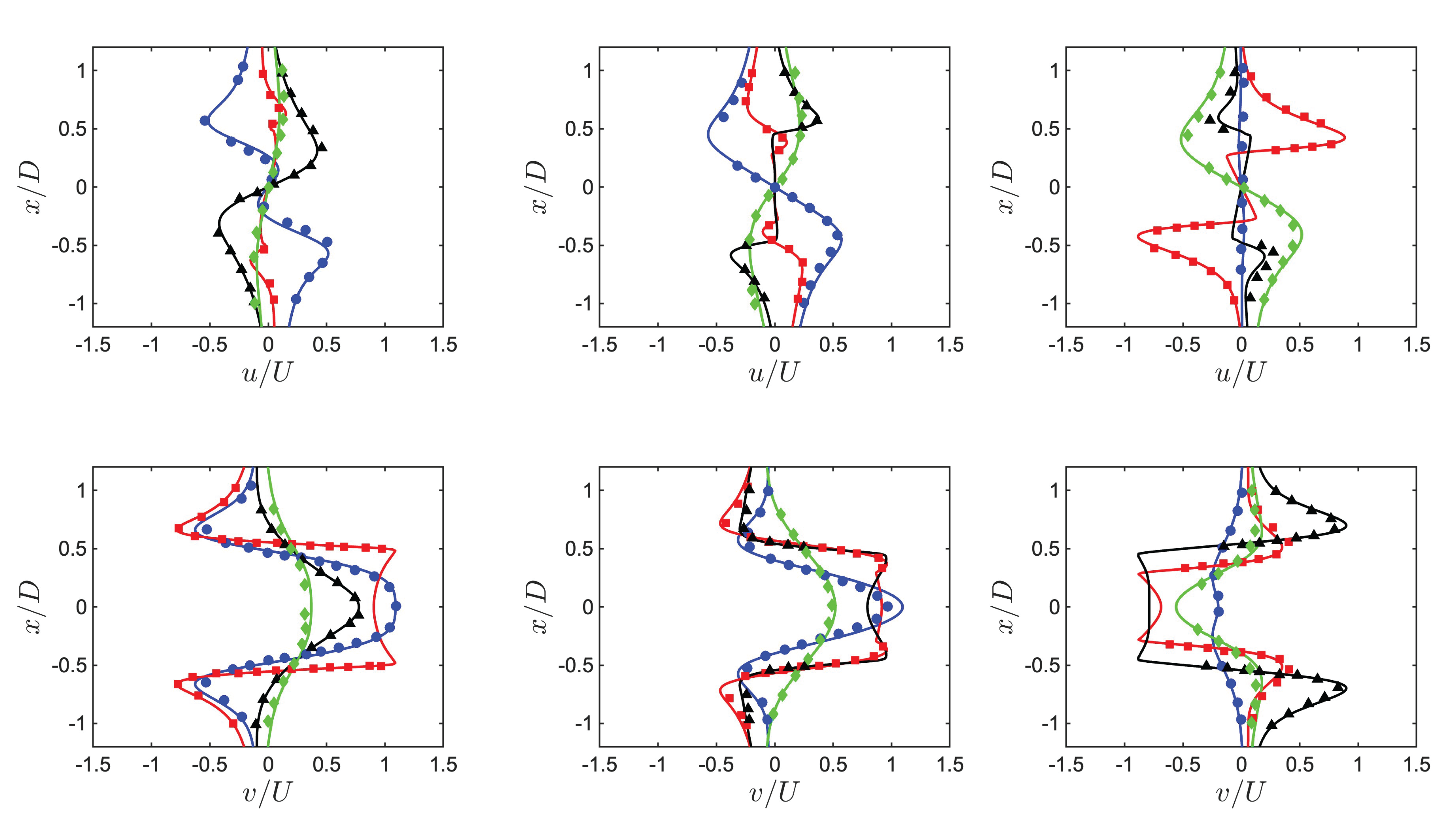}}
  \caption{Velocity profiles for three different phase angles (a) $\phi=180^o$, (b) $\phi=210^o$, (c) $\phi=330^o$ for the case $Re=100$ and $KC=5$. Lines are the present numerical results, and symbols are the experimental results by D{\"u}tsch et al.\cite{Dutsch1998JFM}. Data with different symbols indicate different locations. }
  \label{fig:osci_cylinder_profile}
\end{figure}

\begin{figure}[ht]
\centering
\includegraphics[width=0.4\linewidth]{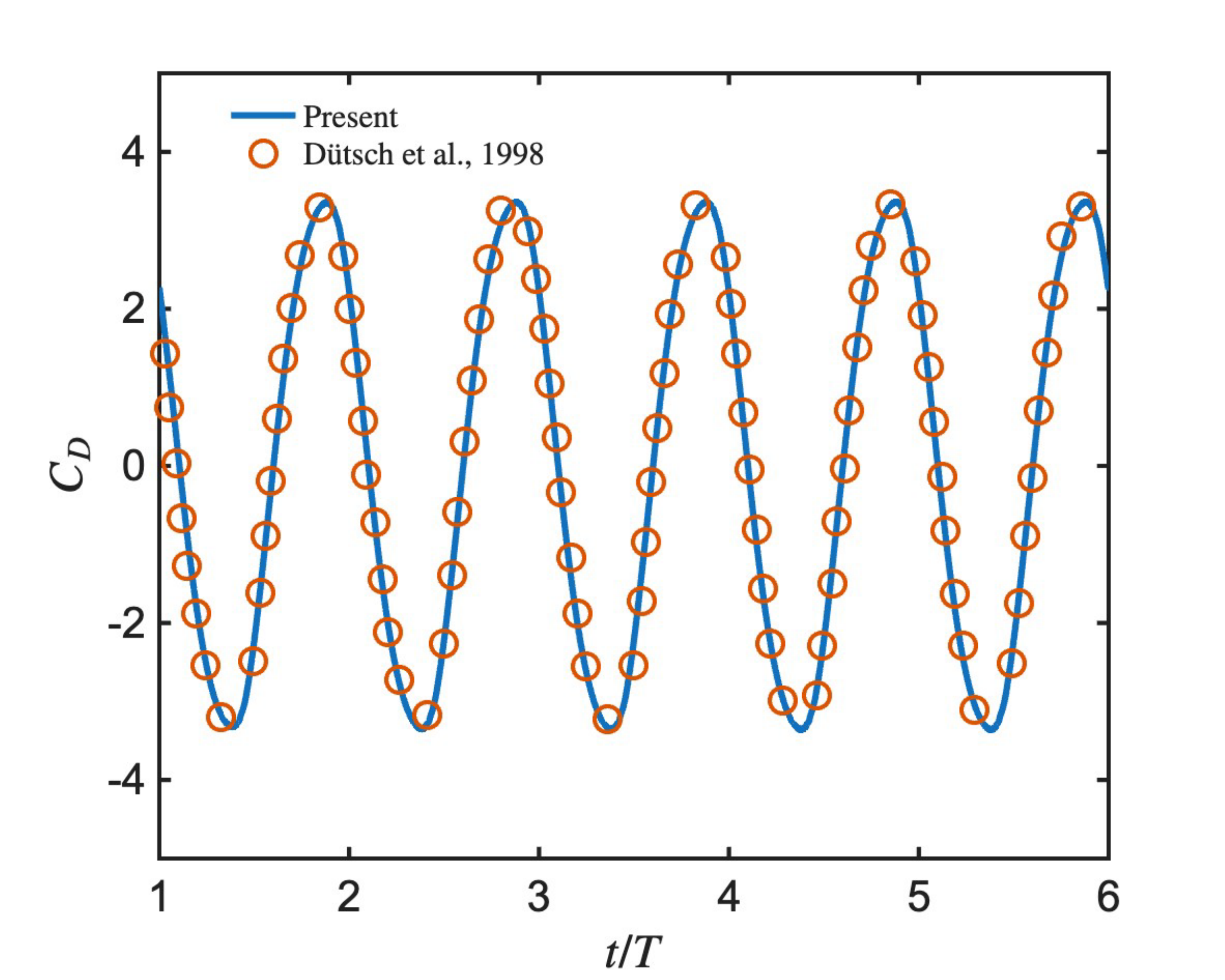}
\caption{Temporal evolution of the drag coefficient ($C_D$) for an oscillating circular cylinder for $Re = 100$ and $KC = 5$. The present results are compared against the numerical reference data of D\"{u}tsch et al.\cite{Dutsch1998JFM}.}
\label{fig:DragCo}
\end{figure}

\section{Validations} \label{sec:validation}
In this section, we perform a series of simulations involving fixed and moving rigid particles in both two- and three-dimensional flows to verify the spatial convergence, accuracy, and robustness of the present method. The validation is carried out by comparing the numerical results with reference solutions available in the literature.

\subsection{2D Oscillating Cylinder in a Quiescent Fluid}
To evaluate the method's accuracy in handling moving boundaries with unsteady wakes, we consider the canonical problem of a circular cylinder oscillating in a stationary fluid. This benchmark has been studied experimentally and numerically by Dütsch et al. \cite{Dutsch1998JFM} and serves as a standard validation case for IBM implementations~\cite{suzuki_effect_2011,cheng2022immersed}.
The cylinder, with diameter $D$, is initially centered in the domain and undergoes harmonic oscillation in the transverse direction. The velocity of the cylinder center, ($U_c,~V_c$), is prescribed as:
\begin{eqnarray}
    U_c(t) &=& -U \cos\left(\frac{2\pi t}{T}\right), \\
    V_c(t) &=& 0,
\end{eqnarray}
where \( U \) is the velocity amplitude and \( T \) is the oscillation period. The flow is governed by the Reynolds number, \( Re = U D / \nu \), and the Keulegan–Carpenter number, \( KC = U T / D \). The grid hierarchy ranges from a base level $l_{min}=7$ to a finest level $l_{max}=12$. At the finest level, the cylinder diameter is resolved with $D=80\Delta x_{l_{max}}$.

A hybrid dynamic refinement strategy is employed to optimize computational resources:
(i) Distance-based refinement: To ensure high fidelity near the moving interface, the grid refinement level is strictly controlled by the distance to the cylinder surface. The finest resolution $l_{max}$ is enforced within a specified envelope around the Lagrangian markers $\mathbf{X}_k$. Beyond this region, the grid level gradually decreases (coarsens) as the distance from the boundary increases.
(ii) Feature-based refinement: In the bulk flow, particularly in the wake region, grid adaptation is driven by the local flow gradients. Cells are tagged for refinement if the normalized velocity gradient exceeds a predefined threshold, ensuring that the shedding vortices are captured accurately even as they convect away from the body.

The cylinder surface is represented by 357 Lagrangian markers. Additionally, internal Lagrangian points are distributed within the cylinder's volume. These internal markers are utilized to compute the rate of change of momentum of the fluid enclosed by the immersed boundary, which is essential for the force coupling calculation in the present method.

In the present study, we adopt the parameter set \( (Re = 100, \, KC = 5) \), following Dütsch et al.~\cite{Dutsch1998JFM}. 
We compare our simulation results with the experimental measurements of Dütsch et al. \cite{Dutsch1998JFM}. 
Figure~\ref{fig:osci_cylinder_profile} shows the velocity profiles \( u \) and \( v \) at three distinct phases ($\phi=180^o,\,210^o,\,$ and $330^o$, where $\phi=2\pi t/T$). The numerical predictions (solid lines) show excellent agreement with the experimental data (symbols) across all phases and locations. The method accurately captures the complex unsteady features, including the flow reversal and the velocity peaks in the wake. 
The temporal evolution of the drag coefficient \( C_D \) is presented in Fig.~\ref{fig:DragCo}, showing that our results closely match those reported by Dütsch et al.~\cite{Dutsch1998JFM}.

\subsection{Settling of a Sphere in a Closed Container}
\label{sec:sphere_settling}
To validate the framework's capability in simulating three-dimensional fluid--structure interactions with significant particle translation, we simulate the sedimentation of a spherical particle in a fluid-filled container. This problem corresponds to the experimental benchmark provided by ten Cate et al.~\cite{Cate2002}, who utilized particle image velocimetry (PIV) to measure the particle trajectory and velocity. 
A sphere of diameter $D_p$ is released from rest in a rectangular container. 
The sphere accelerates under gravity until it approaches the bottom wall, where hydrodynamic lubrication forces cause it to decelerate. The problem is governed by the density ratio $\rho_p/\rho_f$ and the particle Reynolds number $Re=u_{\infty}D_p/\nu$, based on the terminal settling velocity $u_{\infty}$. The simulation parameters for the selected test cases are summarized in Table~\ref{tab:sphere_params}.

To comprehensively assess our framework, the simulations for this problem are conducted in two stages: a baseline validation for physical accuracy, and a sensitivity analysis for computational efficiency. 
For the baseline validation of the fluid-structure interaction, we employ a uniformly refined grid at the maximum resolution level ($l_{max}=9$). At this level, the sphere diameter is resolved with $D_p = 48\Delta x_{l_{max}}$, which provides sufficient resolution to accurately capture the boundary layer dynamics. No-slip boundary conditions are imposed on all container walls.
We note that in the original experiment~\cite{Cate2002}, the top boundary was a free surface. However, following previous numerical studies~\cite{feng2009robust,feng2005proteus}, we model the top surface as a rigid no-slip wall. Given the sufficient distance of the top wall from the sphere's initial position, this approximation has a negligible effect on the settling dynamics.
For the subsequent evaluation of the adaptive mesh algorithm, we utilize a dynamic octree grid with a coarse base level of $l_{min}=7$, allowing dynamic refinement up to $l_{max}=9$. 
In all simulations, we maintained $\Delta_s\in[0.8,1.0]\Delta x_{l_{max}}$ to avoid numerical oscillations, where $\Delta_s$ is the marker spacing. Furthermore, a multi-direct forcing scheme with 5 iterations was employed to enforce the no-slip boundary condition at the particle interface.
In this adaptive setup, refinement is applied dynamically based on two criteria: (i) a geometric envelope ensuring maximum resolution around the sphere interface, and (ii) a flow-based criterion where cells are refined if the normalized local velocity gradient, $|\nabla u| / (u_{\infty}/D_p)$, exceeds a specified threshold $\epsilon_{ref}$. Here, $u_{\infty}$ denotes the reference maximum settling velocity obtained from the uniform fine-grid baseline. 
To maintain a consistent refinement sensitivity across the non-uniform mesh, the applied threshold is dynamically rescaled according to the local grid level.
To avoid numerical instabilities associated with under-resolved lubrication forces at the exact moment of impact, the simulation is terminated when the gap between the particle and the bottom wall falls below $2\Delta x_{l_{max}}$. 

\begin{figure}
\centering
\includegraphics[width=1.0\linewidth]{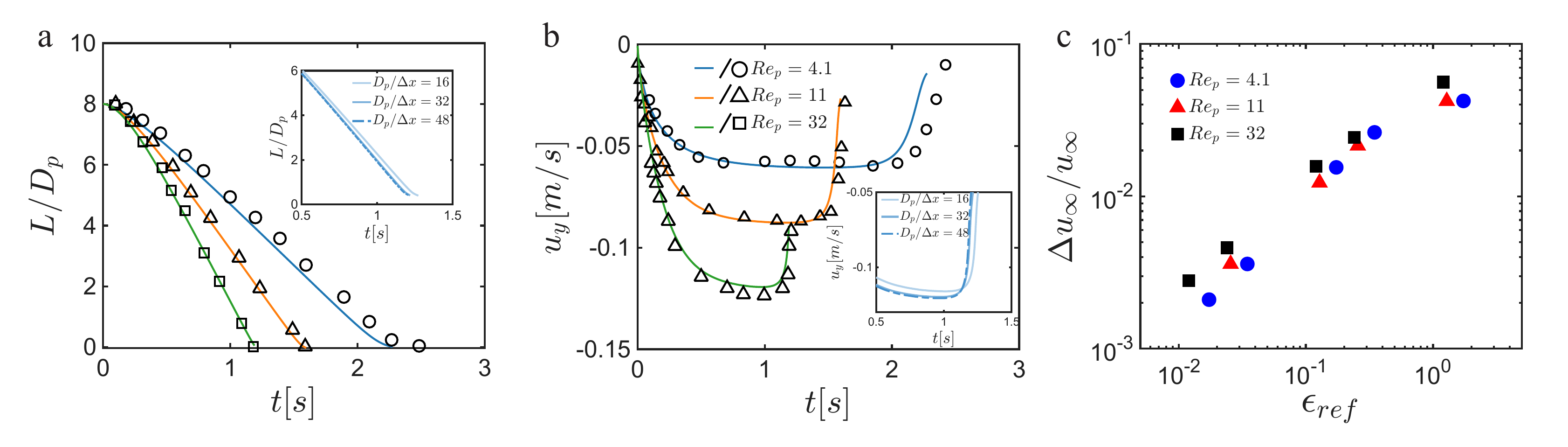}
\caption{Comparison of numerical results (lines) with the experimental data (black open symbols) of ten Cate et al.~\cite{Cate2002} for sphere sedimentation: (a) Time evolution of the particle height (distance to bottom). (b) Time evolution of the vertical settling velocity. The insets in panel (a) and (b) zooms in on the time window $t = 0.5-1.5$\,s to highlight the spatial convergence for the $Re_p=32$ case. (c) Normalized error in terminal velocity as a function of the grid refinement threshold $\epsilon_{ref}$, demonstrating the convergence of the adaptive method to the uniform grid solution.}
\label{fig:sphere_settle}
\end{figure}

Figures~\ref{fig:sphere_settle}(a) and \ref{fig:sphere_settle}(b) compare the time evolution of the settling velocity and the particle's vertical position obtained from simulations with uniform grids against the experimental data of ten Cate et al.~\cite{Cate2002}. The simulation results (solid lines) show excellent agreement with the experiments (symbols) throughout the entire process: the initial acceleration, the steady settling phase, and the deceleration due to the wall effect.
The grid convergence was performed for the $Re_p=32$ case, which requires the highest grid resolution. As shown in the insets of \ref{fig:sphere_settle} (a) and (b), the time evolution of the particle height and settling velocity converges asymptotically as the grid is refined. Quantitatively, the relative errors in terminal velocity are 3.8\% and 0.73\% for $D_p/\Delta x_{l_{max}}=16$ and 32, respectively, when compared to the finest resolution of $D_p/\Delta x_{l_{max}}=48$. This confirms that the resolution of $D_p/\Delta x_{l_{max}}=48$ yields a grid-independent solution. Note that a distance-based refinement strategy, which enforces the finest grid level for the cells within a radial distance of 2$r_p$ from the particle center, was applied to ensure the physical extent of the finest grid remained identical across all resolutions.

Building upon this validated baseline, we then quantify the trade-off between computational cost and accuracy introduced by the adaptive mesh by performing a sensitivity analysis on the refinement threshold $\epsilon_{ref}$. A lower threshold results in a larger refined region (filling more of the wake), while a higher threshold leads to a sparser grid.
Figure~\ref{fig:sphere_settle}(c) illustrates the normalized error $\Delta u_\infty=|u_{max}-u_{\infty}|$ in the maximum terminal velocity as a function of the refinement threshold. The results demonstrate that the adaptive grid method converges to the uniform grid solution as the threshold is reduced, allowing us to achieve high accuracy with a significantly reduced total cell count.

\begin{table}[h!]
\centering
\caption{Physical parameters for the sphere sedimentation cases, corresponding to the experiments of ten Cate et al.~\cite{Cate2002}.}
\label{tab:sphere_params}
\begin{tabular}{ccccc}
\hline
Case & $\rho_p \, (\mathrm{kg/m^3})$ & $\rho_f \, (\mathrm{kg/m^3})$ & $\nu \, (\mathrm{m^2/s})$ & $Re$ \\
\hline
1 & 1120 & 965 & $2.2\times10^{-4}$ & 4.1 \\
2 & 1120 & 962 & $1.17\times10^{-4}$ & 11.6 \\
3 & 1120 & 960 & $6.04\times10^{-5}$ & 31.9 \\
\hline
\end{tabular}
\end{table}

\section{Application: Bubble-Particle Collision Dynamics} \label{sec:app}
In this section, we apply the proposed computational framework to investigate the hydrodynamics of bubble-particle collisions, a fundamental process governing the capture efficiency in froth flotation.

In typical flotation cells, the bubble Weber number  is generally low, indicating that surface tension dominates over inertial forces, thereby maintaining the bubble's spherical shape. Furthermore, industrial systems invariably contain surface-active agents (surfactants) and fine particles that accumulate at the interface. This contamination immobilizes the gas-liquid interface, rendering it essentially rigid with a no-slip boundary condition~\cite{chan_bubbleparticle_2023, chan2026predictive,Nguyen2004}.
Consequently, it is physically justified to model the bubbles as finite-size rigid spheres.
This problem poses a unique multi-scale challenge: the bubble diameter ($d_b$) is typically much larger than the dissipative scales of the flow, requiring fully resolved fluid-structure coupling (IBM), whereas the mineral particles ($d_p$) are orders of magnitude smaller and are commonly treated as Lagrangian point masses. Our LBM-IBM-AMR framework is well-suited to overcome these inherent simulation challenges.

\subsection{Bubble-Particle Collision in Quiescent Flow}
We first consider the collision between a rising bubble and inertialess particles in a quiescent fluid. This deterministic case serves as a critical benchmark for validating the framework's capability to capture the fine-scale hydrodynamic interactions (specifically, the interception mechanism) in the boundary layer.

We simulate a spherical bubble of diameter $d_b$ rising at a constant terminal velocity $V_b$. To facilitate long-time integration, we employ a reference frame attached to the bubble: the bubble is fixed at the center of the domain, and a uniform downward flow $U_\infty=V_b$ is imposed at the inlet.
The flow regime is characterized by the bubble Reynolds number, $Re_b=V_bd_b/\nu$. We focus on the range $20\leq Re_b\leq 150$, where the wake remains steady and axisymmetric.
The particles are seeded near the inlet only after the flow has reached a steady state. 
The particles are seeded uniformly and randomly within a circular region, which corresponds to the projection of the bubble. Since the collision efficiency is inherently lower than 1, particles starting outside this projected stream tube will hydrodynamically bypass the bubble. The initial velocities of the particles are set equal to the local fluid velocity at the inlet. In this scenario, releasing a number of $O(10^4)$ particles for each particle radius is sufficient to achieve a high local particle number density and ensure the statistical convergence of the calculated collision efficiency. An open boundary condition is applied at the outlet, meaning particles are simply removed from the computation once they exit the domain.
The particles are assumed to be weakly-inertial (Stokes number $St_p\ll 1$), meaning they follow the fluid Lagrangian trajectories. Consequently, collisions in this regime are exclusively driven by the interception mechanism, meaning a collision occurs when the finite volume of the small particle geometrically overlaps with the bubble.
Therefore, a collision event is registered and counted whenever the center-to-center distance between a particle and the bubble falls below their combined radii, i.e., $|\mathbf{x}_b - \mathbf{x}_p|\leq r_p+r_b$.
During the simulation, collision detection is performed at every time step. Once a particle collides with the bubble, it is counted and subsequently removed from the simulation to avoid repeated counting.

A major computational challenge in this problem is the conflicting requirement for domain size and resolution.
On one hand, the domain size $L$ must be sufficiently large to minimize blockage effects, which significantly alter the drag and collision dynamics at low Reynolds numbers. Magnaudet et al. \cite{Magnaudet1995} suggested $L\geq40r_b$ for accurate results.
On the other hand, accurately resolving the thin boundary layer at the bubble surface requires a very fine mesh, with the resolution requirement scaling as $\simeq Re_b^{1/2}$\cite{johnson_flow_1999}.
Discretizing such a large domain uniformly with the fine resolution required at the interface would be computationally prohibitive.
Here, the adaptive grid strategy is indispensable. We set the domain size to a cubic box of length $L=40.96d_b$. 
The grid is dynamically refined near the bubble surface (level $l_{max}=12$) and coarsened in the far field ($l_{min}=8$).
The dynamic adaptation in the bulk flow is driven by the normalized velocity gradient. We employ a refinement threshold of $\epsilon_{ref}=0.1$. 
This specific threshold is selected based on the sensitivity analysis performed in the sphere sedimentation validation (Section~\ref{sec:sphere_settling}), where it was demonstrated to yield a relative error of merely 1\% compared to the uniform fine-grid benchmark.
This configuration yields a resolution of approximately 50 grid points per bubble radius ($r_b/\Delta_{x}=50$) at the interface, ensuring high-fidelity capture of the interception trajectories, while maintaining computational efficiency.
The computational advantage is evident in the total grid count. The adaptive simulation peaked at $4\times10^6$ grid points, whereas an equivalent uniform grid would require $6.9\times10^{10}$ grid points. 
This corresponds to a massive reduction in memory requirements, with the adaptive grid using merely 0.006\% of the resources of a uniform approach.

\begin{figure}[t]
\centering
\includegraphics[width=1.0\linewidth]{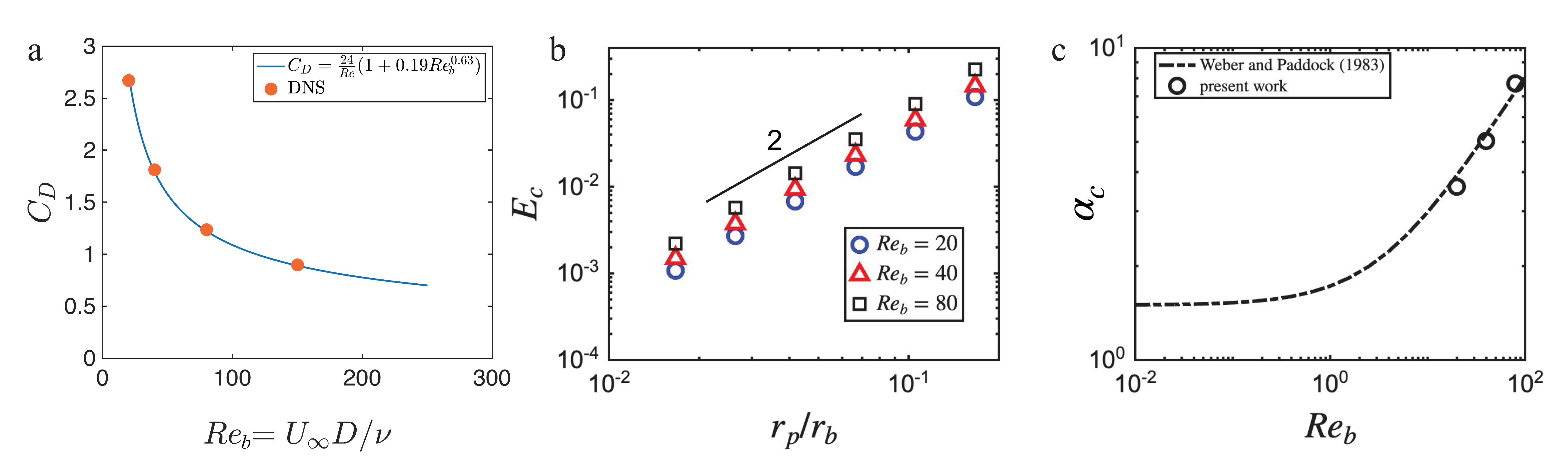}
\caption{(a) Drag coefficient $C_D$ of the bubble as a function of Reynolds number $Re_b$. The solid line represents the standard empirical correlation~\cite{Schiller1933}. (b) Collision efficiency $E_c$ versus particle-to-bubble radius ratio $r_p/r_b$ for different $Re_b$, demonstrating the square-law scaling. (c) The proportionality coefficient $\alpha_c$, extracted from the power-law fits $E_c=\alpha_c(r_p/r_b)^2$, as a function of $Re_b$. Open circles denote present simulation results; the dashed line indicates the theoretical prediction by Weber and Paddock~\cite{weber_interceptional_1983}.}
\label{fig:bupar_stillflow}
\end{figure}

We first validate the flow field by examining the drag coefficient for the bubble, which is defined as $C_D=2F_d/(\rho_fU_\infty^2\pi r_b^2)$ ($F_d$ is the drag force). 
Figure~\ref{fig:bupar_stillflow}(a) compares the computed $C_D$ against the standard empirical correlation for rigid spheres~\cite{Schiller1933}. The simulation results show excellent agreement with the empirical curve, confirming that the adaptive mesh accurately resolves the velocity gradients at the bubble surface and the near-wake structure.

The collision efficiency, $E_c$, is defined as the ratio of the number of particles that actually collide with the bubble to the number of particles flowing through the projected cross-sectional area of the bubble.
It is important to emphasize that the current model focuses purely on hydrodynamics and neglects film drainage and surface forces. Therefore, the 'collision efficiency' evaluated in this study strictly represents the hydrodynamic collision probability. It should be distinguished from the 'attachment efficiency' or 'capture efficiency', which would require additional modeling of physical-chemical attachment and detachment processes.
For inertialess particles, the collision is driven by the interception mechanism.
At the no-slip surface, the fluid velocity vanishes, and the velocity profile grows linearly with distance from the wall. Consequently, theoretical analysis for Stokes flow predicts that $E_c$ scales with the square of the size ratio, i.e., $E_c\propto(r_p/r_b)^2$ \cite{Nguyen2004}.
For finite Reynolds numbers, Weber \& Paddock \cite{weber_interceptional_1983} proposed a correction to account for the boundary layer thickness:
\begin{equation}
E_c = \frac{3}{2} \left(\frac{r_p}{r_b}\right)^2 \left( 1 + \frac{3/16 Re_b}{1 + 0.249 Re_b^{0.56}} \right).
\label{eq:weber_paddock}
\end{equation}
Figure~\ref{fig:bupar_stillflow}(b) plots the computed collision efficiency as a function of the size ratio $r_p/r_b$ for various $Re_b$. 
A clear power-law scaling with an exponent of 2 is observed , consistent with the theoretical interception mechanism.
To quantitatively evaluate the influence of the bubble Reynolds number, we fit the simulation data for each $Re_b$ using the relation $E_c=\alpha_c(r_p/r_b)^2$.
The extracted proportionality coefficient, $\alpha_c$ (which effectively represents the normalized collision efficiency $E_c/(r_p/r_b)^2$), is plotted as a function of $Re_b$ in Figure~\ref{fig:bupar_stillflow}(c).
Our results (open circles) show good agreement with the theoretical prediction (dashed line) and are consistent with previous body-fitted mesh simulations~\cite{sarrot_determination_2005}.
This validation demonstrates that our Cartesian-grid-based LBM-IBM framework, augmented by adaptive refinement, achieves accuracy comparable to body-fitted methods for delicate near-wall particle tracking problems.

\begin{figure}[t]
\centering
\includegraphics[width=1.0\linewidth]{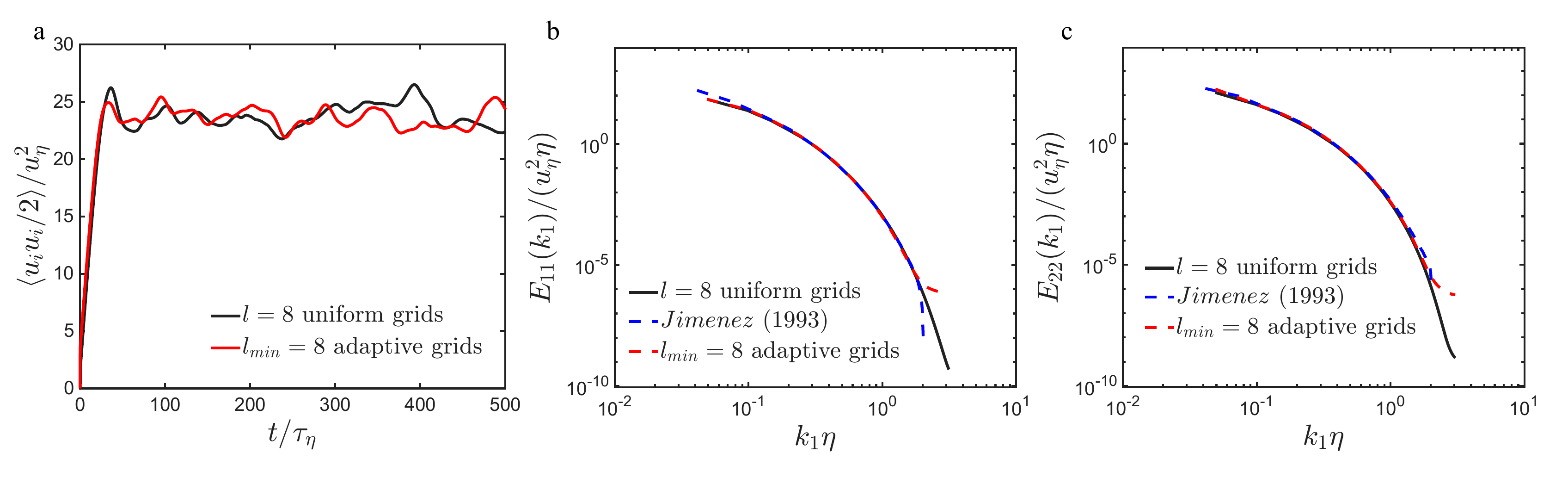}
\caption{(a) Temporal evolution of the normalized turbulent kinetic energy comparing the adaptive grid simulation (with a moving refinement zone) and a uniform grid benchmark. (b) Longitudinal and (c) Transverse energy spectra. The adaptive mesh results (solid lines) closely match the uniform mesh results (red dashed-lines) and the reference literature (blue dashed-lines).}
\label{fig:hitspectrum}
\end{figure}

\begin{figure}[t]
\centering
\includegraphics[width=0.8\linewidth]{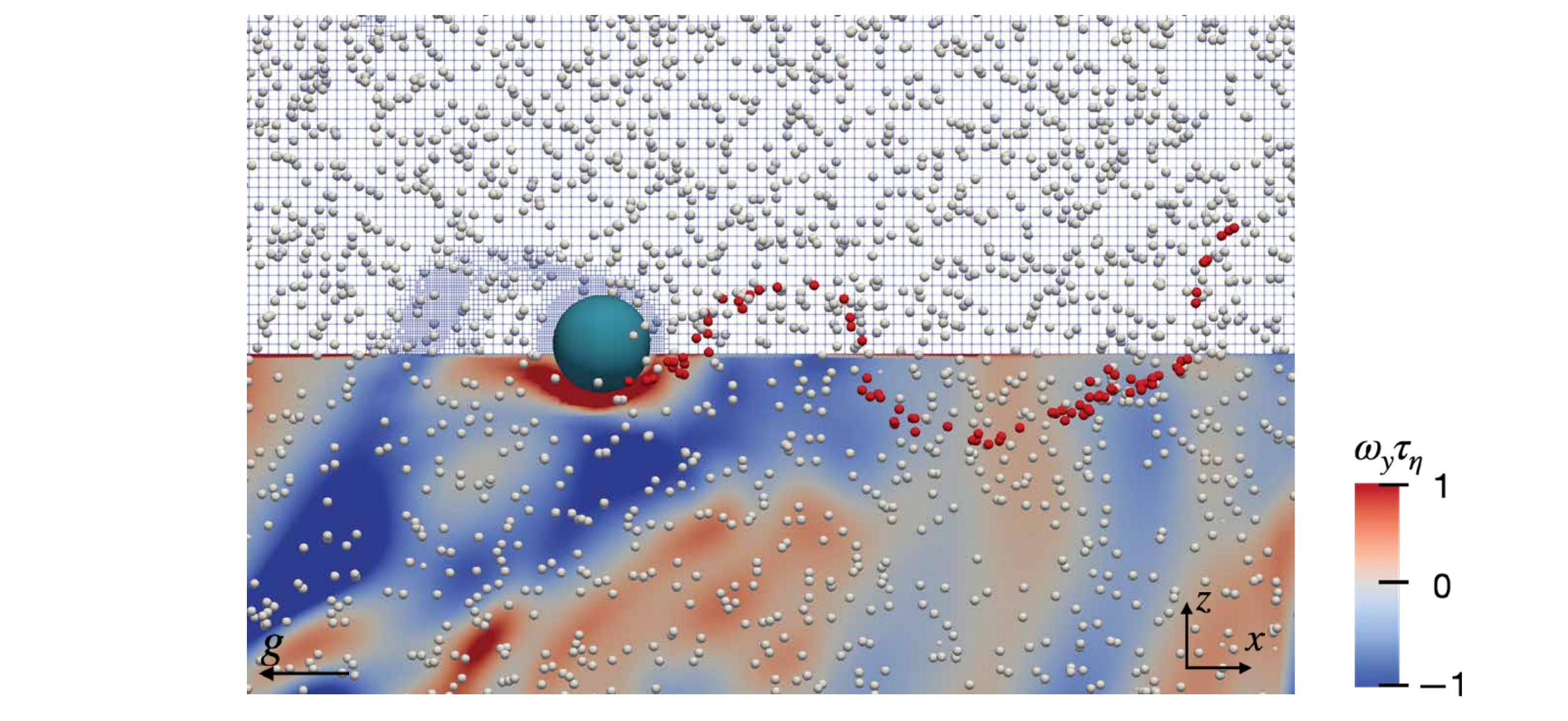}
\caption{Snapshot of the flow slice across the bubble center, with gravity pointing to the left. The upper half panel visualizes the instantaneous computational mesh, while the lower half displays the $y$-component of the vorticity ($\omega_y$). Particles that are destined to collide with the bubble during the simulation are shown in red.}
\label{fig:hitsnapshot}
\end{figure}
\subsection{Bubble-Particle Collision in Isotropic Turbulence}
\label{sec:hit_collision}

We now extend our framework to the simulation of bubble-particle interactions in homogeneous isotropic turbulence (HIT).
The background turbulent flow is generated and sustained using a large-scale linear forcing scheme~\cite{Perlekar2012}. This method injects energy into the low wavenumbers at a rate that balances the viscous dissipation, ensuring that the resulting turbulence is statistically stationary, homogeneous, and isotropic.
The flow is characterized by the Taylor microscale Reynolds number, $Re_\lambda = u'\lambda/\nu$, where $u'$ is the single-component root-mean-square (r.m.s.) fluid velocity and $\lambda=\sqrt{15\nu u'^2/\varepsilon}$ is the Taylor microscale, with $\varepsilon$ being the mean energy dissipation rate.

Before introducing the dispersed phases, it is imperative to verify that the dynamic grid adaptation does not introduce spurious artifacts or degrade the spectral properties of the turbulence, i.e., specifically the transfer of energy across refinement boundaries.
To this end, we perform a precursor simulation in a cubic domain with periodic boundary conditions.
Instead of a physical bubble, we impose a "phantom" spherical refinement region with a radius of $10\eta$ (where $\eta$ is the Kolmogorov length scale) that translates through the domain. This forces the grid to dynamically refine to level $l_{max}=10$ inside the sphere and coarsen to $l_{min}=8$ outside, mimicking the mesh topology of a moving bubble simulation.
The target Reynolds number is set to $Re_\lambda=64$.

Figure~\ref{fig:hitspectrum}(a) compares the time evolution of the normalized mean kinetic energy, $\langle u_iu_i/2\rangle/u_\eta^2$, obtained from the adaptive grid simulation against a benchmark simulation using a uniform static grid at level $l=8$.
The adaptive case maintains the same stationary energy level as the uniform case, confirming that the dynamic refinement process is energy-conservative.
Furthermore, the longitudinal and transverse energy spectra, shown in Figs.\ref{fig:hitspectrum}(b) and (c), exhibit excellent agreement between the adaptive and uniform grid results. Both spectra correctly reproduce the dissipative range scaling consistent with literature data \cite{jimenez1993structure}.
This validation confirms that our adaptive framework can sustain realistic turbulent fluctuations without introducing numerical dissipation or aliasing errors at grid interfaces.

Having validated the carrier phase, we proceed to the full application of bubble-particle collisions. 
We seed 20 fully contaminated bubbles (corresponding to a volume fraction of 
0.5 \%) and $10^6$ inertial point particles into the developed turbulence at $Re_\lambda=64$. 
Note that the $10^6$ point particles serve as non-interacting statistical probes. Due to the one-way coupling, their independent trajectories act as an ensemble average, which accelerates the statistical convergence of rare collision events without compromising the dilute assumption.
The size ratio between the particles and bubbles is $r_p/r_b=0.1$. The bubbles are resolved with 40 grid points per diameter ($d_b/\eta=5$) at the maximum refinement level $l_{max}=10$, while the background turbulence is captured at $l_{min}=8$.
It should be noted that a full-scale production case typically involves higher $Re_\lambda$ and domain sizes, demanding higher computational resources. Consequently, the current setup is designed as a scaled-down, representative test case. Despite the reduced parameters, this configuration retains all the essential multiscale complexities of the physical process, serving to strictly demonstrate the framework's coupled capabilities. 
Because the simulation domain is fully periodic in all directions, the non-zero net vertical force (buoyancy) acting on the rising bubbles would induce a continuous, unphysical acceleration of the bulk fluid. To prevent this artificial mean flow and ensure the turbulence remains statistically stationary, we require the net force exerted by the dispersed phase on the fluid to be zero on average~\cite{Hoefler2000, chouippe_forcing_2015}. This is achieved by subtracting the instantaneous spatial average of the IBM forcing field from the local forcing at every time step:
\begin{equation}
\mathbf{F}(\mathbf{x},t) = \mathbf{F}^{ibm}(\mathbf{x},t) - \langle \mathbf{F}^{ibm} \rangle_\Pi(t), \label{eq:compens_ibm}
\end{equation}
where $\langle \mathbf{F}^{ibm}\rangle_\Pi(t) = \frac{1}{||V_\Pi||}\int \mathbf{F}^{ibm}(\mathbf{x},~t) \rm{d}\mathbf{x}$ represents the mean force density over the entire domain volume $V_\Pi$. 
It should be noted that the particles and bubbles are not neutrally buoyant. The density ratio of the large fully contaminated bubbles to the fluid is set to $\rho_b/\rho_f=0.001$, while for the small particles, it is $\rho_p/\rho_f=5$. To overcome the added-mass instability induced by such large density disparities for the lightweight bubbles, the rigid body motion is solved using the robust semi-implicit coupling scheme described by Ref.\cite{Tschisgale2017}.

With this numerical configuration, we evaluate the interaction dynamics. Figure~\ref{fig:hitsnapshot} presents a cross-sectional snapshot through a bubble center, with the gravity vector pointing to the left. We explicitly mark the particles that eventually collide with the bubble in the subsequent time steps in red. It is clearly observed that the turbulent fluctuations sweep the particles and the colliding particles come from a broad, irregularly shaped upstream region ahead of the bubble,  compared to the deterministic trajectories in quiescent flow conditions.

\begin{figure}[t]
\centering
\includegraphics[width=1.0\linewidth]{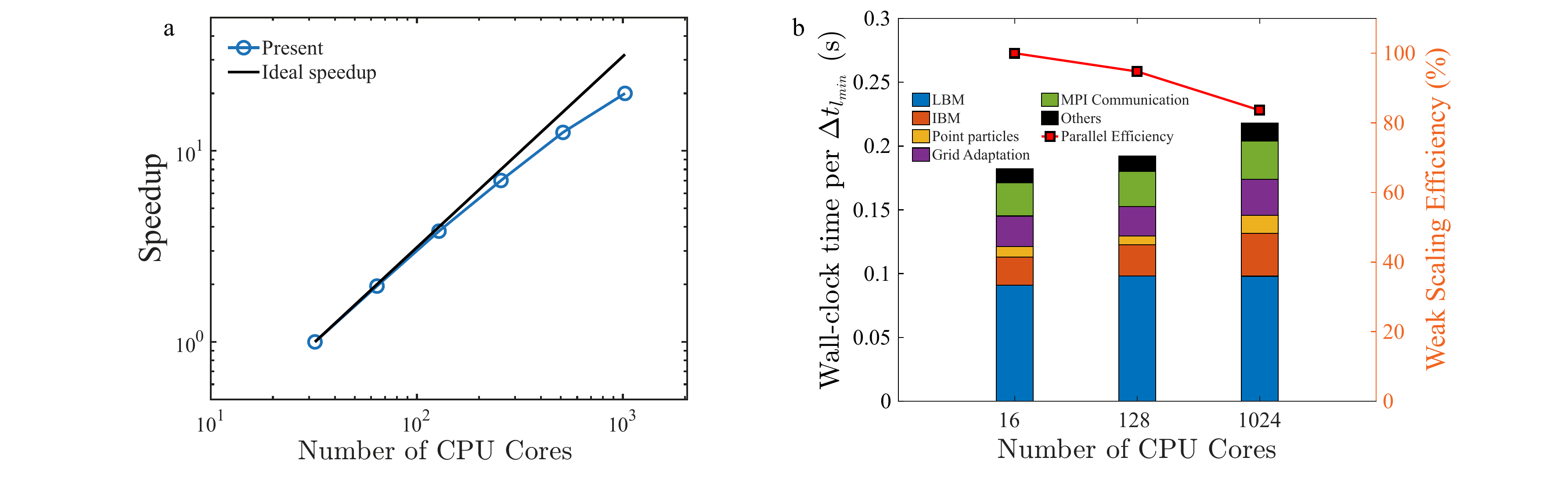}
\caption{(a) Strong scaling performance of the turbulent application. Here, the speedup is evaluated relative to the 32-processor baseline, defined as $Speedup(N)=t(32)/t(N)$, where $t(N)$ is the computational time using N processors. The solid line represents the ideal speedup. (b) Weak scaling performance from 16 to 1024 cores. To maintain a constant workload per core, the domain volume and the number of resolved bubbles are scaled proportionally (from 1 bubble on 16 cores to 64 bubbles on 1024 cores). The stacked bar chart (left axis) displays the absolute wall-clock time per step broken down by major algorithmic components, while the dotted line (right axis) represents the overall weak scaling efficiency.}
\label{fig:performance}
\end{figure}

\subsection{Computational Performance and Scalability}
To demonstrate the parallel scalability of the current framework, we evaluate both its strong and weak scaling performances. Note that strong scaling measures the speedup for a fixed total problem size, whereas weak scaling assesses the efficiency when the problem size scales proportionally with the number of processors. First, Figure \ref{fig:performance}(a) evaluates the strong scaling performance of the full large-scale application (described in Section \ref{sec:hit_collision}) up to 1024 processors. As the number of processors increases, the computational load becomes increasingly heterogeneous, given that the intensive IBM operations and dense local grids are strictly tied to the dynamic bubble locations. Despite this inherent dynamic load imbalance, the continuous re-partitioning of the \texttt{p4est} octree ensures that the overall speedup remains robust, deviating moderately from the ideal linear scaling curve at the highest core counts. 

To further substantiate the framework's capability and provide a detailed computational cost breakdown, a systematic weak scaling test was conducted scaling from 16 up to 1024 CPU cores. To maintain a constant computational workload per core, the problem size was scaled proportionally with the number of MPI ranks. Specifically, the baseline case on 16 cores simulated a single finite-size bubble ($r_b/\Delta x_{l_{max}}=40$) within a domain of refinement levels $l_{min}=7$ to $l_{max}=9$, where $5\times 10^4$ point-like particles are seeded uniformly in space. For the 128-core and 1024-core cases, the domain volume was increased by factors of 8 and 64, by raising the grid levels to ($l_{min}=8$, $l_{max}=10$) and ($l_{min}=9$, $l_{max}=11$), respectively. Correspondingly, the number of resolved bubbles was increased to 8 and 64, while the number of point-like particles was scaled proportionally to $4\times10^5$ and $3.2\times10^6$. A fixed distance-based refinement criterion ($< 2r_b$) was employed to ensure consistent local load distribution. 
Figure \ref{fig:performance}(b) shows the weak scaling results, illustrated by a stacked bar chart of the wall-clock time per step $\Delta t_{l_{min}}$ divided by algorithmic components along the left-axis. The parallel efficiency is indicated by the right-axis of the figure. The framework demonstrates excellent scalability with a weak scaling efficiency of 94\% at 128 cores and 84\% at 1024 cores. 
The profiling breakdown reveals that the core computational routines consume a nearly-constant wall-clock time across all core counts. This confirms that the host-cell search algorithm and the auxiliary grids effectively prevent Lagrangian tracking from becoming a parallel bottleneck. 
It is worth noting that a slight increase in the wall-clock time for the IBM operations is observed at higher core counts. This occurs because the IBM routine includes the rigid-body time-stepping of the finite-size bubbles. Integrating the equations of motion requires accumulating hydrodynamic forces and torques distributed across the MPI context, which necessitates global reduction operations (e.g., {$MPI\_Allreduce$}). Consequently, as both the number of bubbles and MPI ranks increase, the communication overhead inherent in these global reductions naturally causes an increase in the IBM computational time.
Overall, these strong and weak scaling results confirm that the present methodology is scalable and suited for massive multi-scale particle-laden flow simulations.

\section{Conclusions}
\label{sec:conclusion}
In this work, we have presented and validated a scalable computational methodology designed for the simulation of multiphase flows involving dispersed phases with large scale disparities. The computational framework couples the lattice Boltzmann method for the fluid dynamics, the Lagrangian particle tracking for point-like advected particles, and the immersed boundary method for finite-size moving bodies, all integrated within a dynamically adaptive octree grid structure managed by the \texttt{p4est} library.

The primary objective was to overcome the computational bottlenecks associated with resolving thin boundary layers around large fast-rising or settling particles (e.g., bubbles) while simultaneously tracking massive numbers of microscopic inertial particles in a turbulent background flow. The key contributions and findings of this study are summarized as follows:
First, we implemented a robust integrated methodology between the LBM and adaptive mesh refinement. By employing a volumetric, cell-centered data structure with recursive time-stepping, we ensured strict conservation of mass and momentum across coarse-fine grid interfaces. The local grid refinement is based on flow gradients, concentrating computational resources solely where high resolution is required.
Second, to address the challenge of Lagrangian particle tracking on distributed, unstructured grids, we used an efficient parallel algorithm based on Space-Filling Curves. By using the Morton ordering of the octree, the host-cell search algorithm achieves a logarithmic complexity of $O(N_p\log N_{cell})$, significantly outperforming bare geometric searches. 
Third, the accuracy and fidelity of the framework were rigorously validated against canonical benchmarks.
In the case of an oscillating cylinder, the method accurately captured the unsteady forces and wake structures.
In the sphere sedimentation case, the simulated trajectories matched experimental data.
Finally, the framework was applied to the fundamental problem of large-bubble small-particle collisions in flotation dynamics. The simulation of a rising contaminated bubble interacting with inertialess particles successfully reproduced the theoretical interception mechanism, recovering the expected $E_c\propto(r_p/r_b)^2$ scaling law and matching the theoretical predictions. 
Furthermore, the code was applied to the flow configuration of homogeneous and isotropic turbulence, simulating fully resolved bubbles interacting with $10^6$ inertial point particles.
Most notably, the AMR strategy showcased immense efficiency in this complex scenario: the adaptive grid reduced the required computational resources by two orders of magnitude for the turbulent test case while maintaining efficient parallel performance. 
Moreover, the computational advantage is expected to become more pronounced at higher $Re_\lambda$ and bubble $Re_b$, where the scale separation between the turbulence and the thin boundary layers inherently widens.
In summary, the proposed LBM-IBM-AMR framework offers a powerful tool for investigating complex fluid-structure interactions where scale separation and computational efficiency are limiting factors.
Our purely geometric collision model neglects short-range lubrication forces. Standard lubrication models introduce a contact singularity that requires arbitrary micro-physical parameters (e.g., surface roughness) to regularize. To avoid such parameterizations within our macroscopic framework, we omit these forces. Thus, our results represent an idealized hydrodynamic upper bound, with regularized sub-grid lubrication reserved for future work.

\section{Code and Data Availability}
The underlying numerical framework relies on the open-source \texttt{p4est} library for adaptive grid management. The specific in-house implementations of the LBM-IBM coupling and the parallel Lagrangian tracking algorithms, as well as the data that support the findings of this study, are available from the corresponding author upon reasonable request.

\section{Acknowledgements}
This work was supported by the European Research Council (ERC) under the European Union’s Horizon 2020 research and innovation programme (Grant Agreement No. 950111, BU-PACT). We acknowledge the EuroHPC Joint Undertaking for awarding us access (Project EHPC-REG-2023R03-178) to the EuroHPC supercomputer Discoverer, hosted by Sofia Tech Park, Bulgaria.
%\appendix

%\section{Sample Appendix Section}

\bibliographystyle{elsarticle-num}





% \end{thebibliography}
\end{document}